\PassOptionsToPackage{unicode}{hyperref}
\PassOptionsToPackage{hyphens}{url}
\PassOptionsToPackage{dvipsnames,svgnames,x11names}{xcolor}
\documentclass[
  12pt]{article}

\usepackage{amsmath,amssymb}
\usepackage{iftex}
\usepackage[T1]{fontenc}
\usepackage[utf8]{inputenc}
\usepackage{textcomp} 
\usepackage{lmodern}
\IfFileExists{upquote.sty}{\usepackage{upquote}}{}
\IfFileExists{microtype.sty}{
  \usepackage[]{microtype}
  \UseMicrotypeSet[protrusion]{basicmath} 
}{}
\makeatletter
\@ifundefined{KOMAClassName}{
  \IfFileExists{parskip.sty}{%
    \usepackage{parskip}
  }{
    \setlength{\parindent}{0pt}
    \setlength{\parskip}{6pt plus 2pt minus 1pt}}
}{
  \KOMAoptions{parskip=half}}
\makeatother
\usepackage{xcolor}
\makeatletter
\ifx\paragraph\undefined\else
  \let\oldparagraph\paragraph
  \renewcommand{\paragraph}{
    \@ifstar
      \xxxParagraphStar
      \xxxParagraphNoStar
  }
  \newcommand{\xxxParagraphStar}[1]{\oldparagraph*{#1}\mbox{}}
  \newcommand{\xxxParagraphNoStar}[1]{\oldparagraph{#1}\mbox{}}
\fi
\ifx\subparagraph\undefined\else
  \let\oldsubparagraph\subparagraph
  \renewcommand{\subparagraph}{
    \@ifstar
      \xxxSubParagraphStar
      \xxxSubParagraphNoStar
  }
  \newcommand{\xxxSubParagraphStar}[1]{\oldsubparagraph*{#1}\mbox{}}
  \newcommand{\xxxSubParagraphNoStar}[1]{\oldsubparagraph{#1}\mbox{}}
\fi
\makeatother

\usepackage{longtable,booktabs,array}
\usepackage{calc} 
\usepackage{multirow,siunitx}
\usepackage{etoolbox}
\makeatletter
\patchcmd\longtable{\par}{\if@noskipsec\mbox{}\fi\par}{}{}
\makeatother
\IfFileExists{footnotehyper.sty}{\usepackage{footnotehyper}}{\usepackage{footnote}}
\makesavenoteenv{longtable}
\usepackage{graphicx}
\graphicspath{{./}}
\makeatletter
\def\maxwidth{\ifdim\Gin@nat@width>\linewidth\linewidth\else\Gin@nat@width\fi}
\def\maxheight{\ifdim\Gin@nat@height>\textheight\textheight\else\Gin@nat@height\fi}
\makeatother
\setkeys{Gin}{width=\maxwidth,height=\maxheight,keepaspectratio}
\makeatletter
\def\fps@figure{htbp}
\makeatother

\makeatletter
\@ifpackageloaded{caption}{}{\usepackage{caption}}
\AtBeginDocument{%
\ifdefined\contentsname
  \renewcommand*\contentsname{Table of contents}
\else
  \newcommand\contentsname{Table of contents}
\fi
\ifdefined\listfigurename
  \renewcommand*\listfigurename{List of Figures}
\else
  \newcommand\listfigurename{List of Figures}
\fi
\ifdefined\listtablename
  \renewcommand*\listtablename{List of Tables}
\else
  \newcommand\listtablename{List of Tables}
\fi
\ifdefined\figurename
  \renewcommand*\figurename{Figure}
\else
  \newcommand\figurename{Figure}
\fi
\ifdefined\tablename
  \renewcommand*\tablename{Table}
\else
  \newcommand\tablename{Table}
\fi
}
\@ifpackageloaded{float}{}{\usepackage{float}}
\floatstyle{ruled}
\@ifundefined{c@chapter}{\newfloat{codelisting}{h}{lop}}{\newfloat{codelisting}{h}{lop}[chapter]}
\floatname{codelisting}{Listing}

\makeatother
\makeatletter
\@ifpackageloaded{caption}{}{\usepackage{caption}}
\@ifpackageloaded{subcaption}{}{\usepackage{subcaption}}
\makeatother

\ifLuaTeX
  \usepackage{selnolig}  
\fi
\usepackage[]{natbib}
\usepackage{bookmark}

\IfFileExists{xurl.sty}{\usepackage{xurl}}{} 
\hypersetup{
  pdftitle={Asymptotic Uniform False Discovery Rate Control for Inference of Time-varying Correlations},
  pdfauthor={Bufan Li; Lujia Bai; Weichi Wu},
  pdfkeywords={time-varying correlation, locally stationary time series, high-dimensional inference, multiplier bootstrap, asymptotic uniform false discovery rate},
  colorlinks=true,
  linkcolor={blue},
  filecolor={Maroon},
  citecolor={Blue},
  urlcolor={Blue},
  pdfcreator={LaTeX via pandoc}}

\usepackage{amsthm}

\theoremstyle{remark}
\newtheorem{remark}{Remark} 
\newtheorem{comment}{Comment}

\usepackage{algorithm}
\usepackage{algpseudocode} 

\newcommand{\bXi}{\boldsymbol \Xi}

\newcommand{\bmu}{\boldsymbol \mu}

\newcommand{\bfeta}{\boldsymbol \eta}

\newcommand{\bx}{{\mathbf x}}
\newcommand{\by}{{\mathbf y}}

\newcommand{\bw}{{\mathbf w}}

\newcommand{\bA}{{\bf A}}

\newcommand{\bI}{{\bf I}}

\newcommand{\bX}{{\bf X}}
\newcommand{\bY}{{\bf Y}}
\newcommand{\bZ}{{\bf Z}}

\newcommand{\bQ}{{\bf Q}}
\newcommand{\bJ}{{\bf J}}

\newcommand{\bG}{{\bf G}}

\newcommand{\cA}{{\cal A}}
\newcommand{\cC}{{\cal C}}

\newcommand{\cE}{{\cal E}}
\newcommand{\cI}{{\cal I}}
\newcommand{\cL}{{\cal L}}

\newcommand{\cG}{{\cal G}}

\newcommand{\cT}{{\cal T}}

\newcommand{\cV}{{\cal V}}

\newcommand{\cR}{{\cal R}}

\newcommand{\cH}{{\cal H}}

\newcommand{\cF}{{\cal F}}

\newcommand{\eZ}{\mathbb{Z}}

\newcommand{\eR}{\mathbb{R}}

\newcommand{\eE}{\mathbb{E}}

\newcommand{\eN}{\mathbb{N}}
\newcommand{\eP}{\mathbb{P}}

\newcommand{\var}{\mathrm{Var}}

\newcommand{\0}{{\bf 0}}

\newcommand{\pth}[1]{\left( #1 \right)}
\newcommand{\qth}[1]{\left[ #1 \right]}
\newcommand{\sth}[1]{\left\{ #1 \right\}}
\newcommand{\nth}[1]{\left\| #1 \right\|}
\newcommand{\ath}[1]{\left| #1 \right|}

\newcommand{\mf}{\mathbf}
\newcommand{\bt}{\mathrm{boot}}

\newcommand{\nb}{\lceil n b \rceil}

\newcommand{\fdp}{\mathrm{FDP}}

\definecolor{revisiongreen}{RGB}{0,100,0}
\def\T{{ \mathrm{\scriptscriptstyle T} }}

\newtheorem{assumption}{Assumption}
\newtheorem{theorem}{Theorem}
\newtheorem{lemma}{Lemma}

\usepackage{multirow}

\begin{document}

\title{\bfseries Asymptotic Uniform False Discovery Rate Control for Inference of Time-varying Correlations}
\author{Bufan Li\thanks{
    We gratefully thank Prof.~Huaqing Jin for providing background information and some relevant references on brain EEG data. Weichi Wu is the corresponding author. Bufan Li and Weichi Wu are supported by the NSFC 12271287. Lujia Bai is supported by the Deutsche Forschungsgemeinschaft (DFG, German Research Foundation, TRR 391: Spatio-Temporal Statistics for the Transition of Energy and Transport) under Project 520388526.}\hspace{.2cm}\\
    Department of Statistics and Data Science, Tsinghua University\\
    Lujia Bai \\
    Department of Mathematics, Ruhr University Bochum\\
    Weichi Wu \\
    Department of Statistics and Data Science, Tsinghua University}
\date{}
\maketitle

\bigskip
\begin{abstract}
\noindent

Inference for locally stationary time series is challenging because the associated hypotheses form an uncountable collection over a continuous time interval, making pointwise false discovery rate (FDR) control inadequate for simultaneous statistical guarantees. We introduce a novel asymptotically uniform false discovery rate (AuFDR), defined as the expectation of the $L_r$-norm of the false discovery proportion (FDP) process where $r$ is allowed to diverge, to quantify and control false discoveries uniformly over time. To operationalize AuFDR control, we develop an inferential framework for time-varying correlations in high-dimensional nonstationary time series that allows for non-Gaussianity, nonlinearity and possible jumps in mean functions. The proposed approach combines robust difference-based estimators with a multiplier-bootstrap procedure to construct uniformly valid time-varying $P$-values. Based on these $P$-values, we propose a time-varying Benjamini--Yekutieli procedure for controlling the AuFDR under arbitrary dependence and establish its asymptotic validity. Extensive simulations demonstrate the finite-sample performance of the proposed method in controlling the AuFDR. Applications to EEG  data and financial time-series data illustrate its practical utility.

\end{abstract}

\noindent%
\vfill

\newpage

\section{Introduction}\label{sec-intro}

\begin{figure}[htbp]
    \centering
    \includegraphics[width=0.95\linewidth]{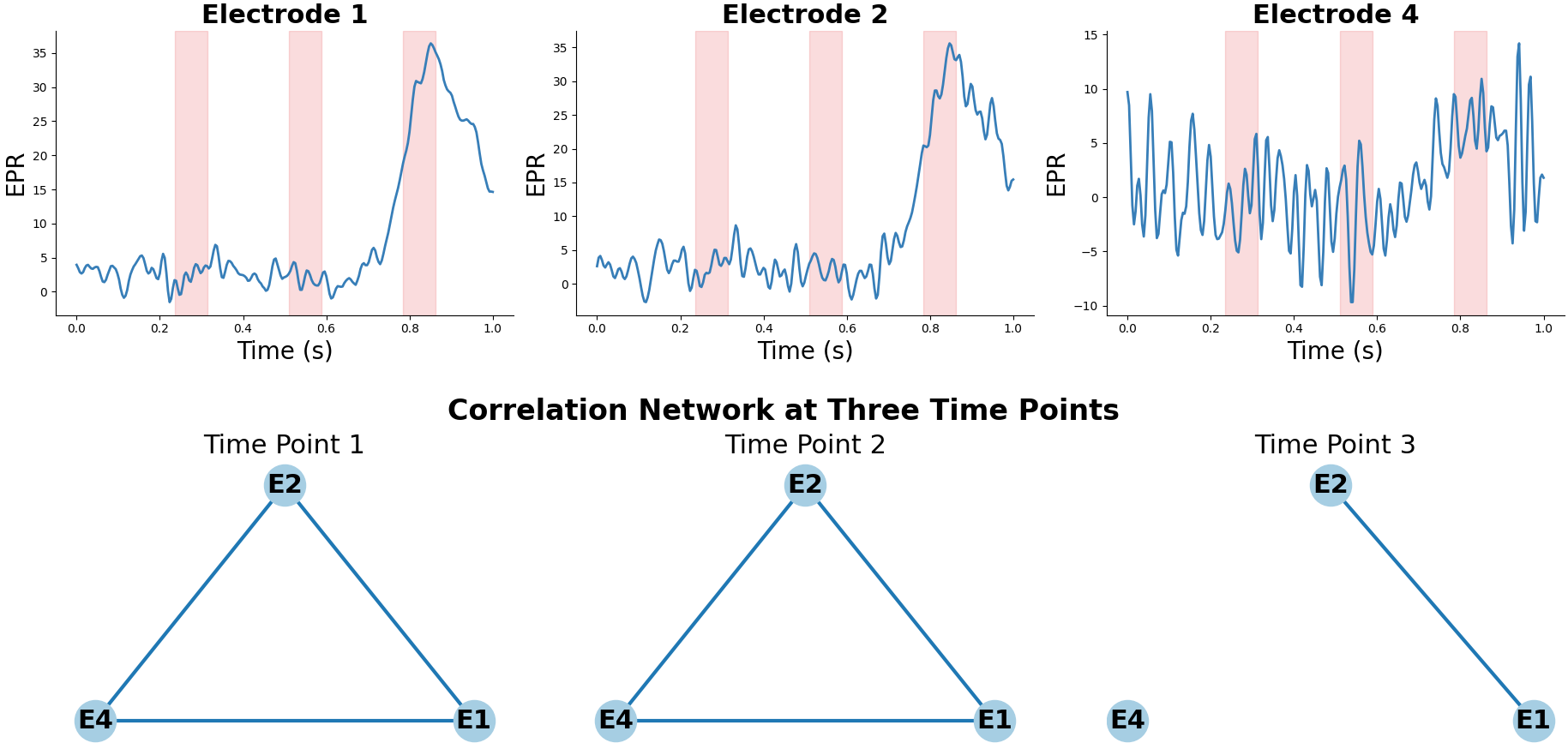}
    \caption{Top: ERP signals from three electrodes, out of $64$ electrodes in total, for a single subject over $n=256$ time points. Red shaded bands denote selected time points. Bottom: Time-varying inference for pairwise correlations visualized as networks, where edges imply the null hypotheses of zero correlation at this time point are rejected.}
    \label{fig:intro}
\end{figure}

A standard approach of error control in large-scale testing is false discovery rate (FDR) control \citep{benjamini2001control}. In the analysis of non-stationary time series under a time-varying scheme, however, additional challenges arise when the hypotheses are indexed by a continuous time parameter, yielding an uncountable family of tests. For instance, in the inference of time-varying correlations (see \cite{hutchison2013dynamic,lurie2020questions} in neuroscience and \cite{onnela2003dynamics,marti2021review} in finance as examples), one is interested in testing the null hypothesis of zero correlation for each pair at each rescaled time point $t \in [0,1]$, yielding a family of hypotheses indexed by both pairs and time. As an example, Figure~\ref{fig:intro} illustrates such an approach on an EEG dataset \citep{zhang1997electrophysiological}.

To conduct reliable inference on non-stationary time series under a time-varying scheme, one needs error control that accounts for the entire trajectory of the time-varying false discovery proportion, \(\mathrm{FDP}(t)\). Conventional pointwise FDR procedures control 
\(\mathrm{FDR}(t)=\mathbb{E}\{\mathrm{FDP}(t)\}\) only at each fixed time point \(t\). As illustrated by the simulation results in Section~\ref{Sec:Base}, such pointwise control methods cannot guarantee that the time-varying FDPs are low globally. Therefore, pointwise error control does not provide global inferential validity for multiple testing in a time-varying framework, and a uniform error criterion is therefore needed.

\begin{figure}
    \centering
    \includegraphics[width=0.8\linewidth]{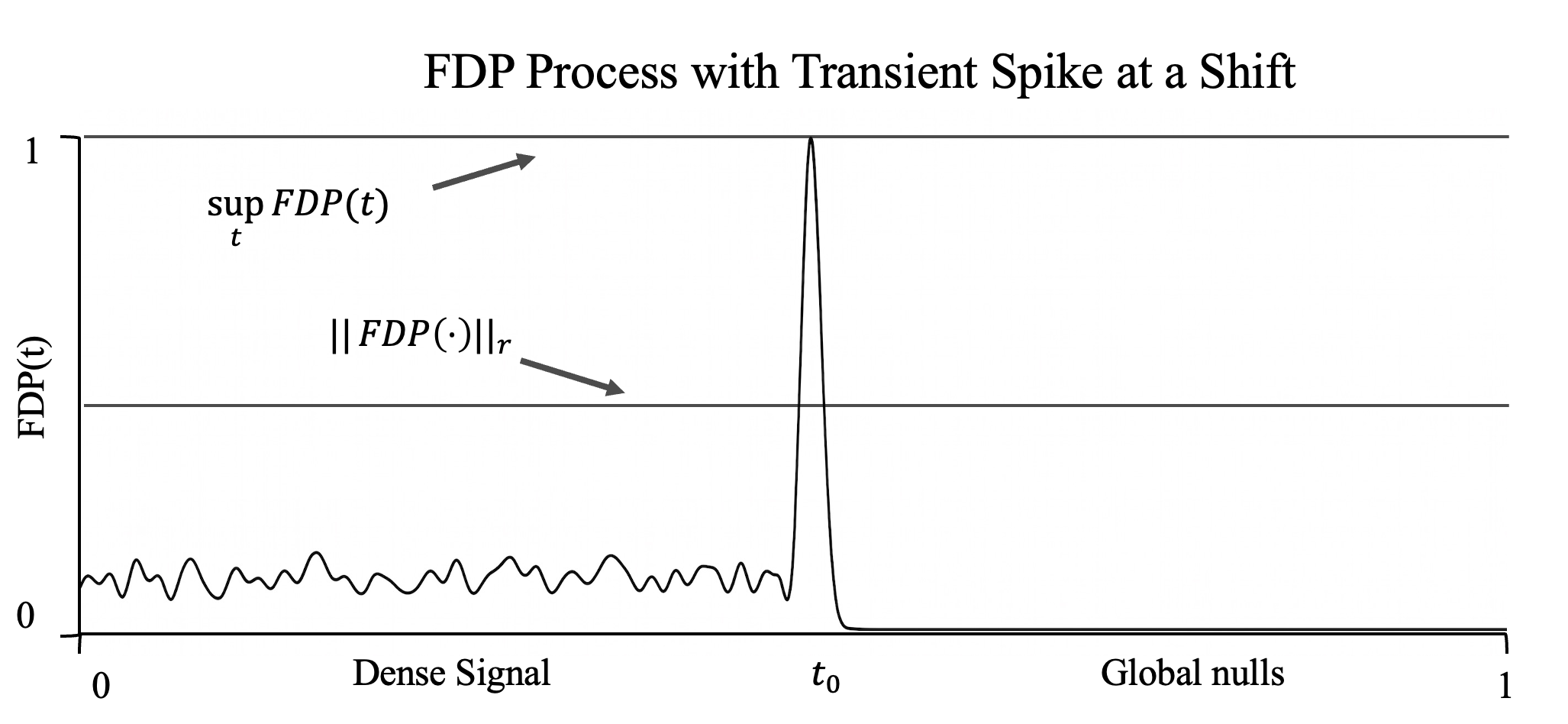}
    \caption{Comparison between $\sup_t \fdp(t)$ and $\|\fdp(\cdot)\|_r$ under time-varying signals.}
    \label{fig:supvsq}
\end{figure}

The supremum-based criterion $\mathbb{E}\{\sup_{t\in[0,1]}\mathrm{FDP}(t)\}$  is a natural worst-case benchmark, but it is not always well aligned with time-varying inference based on local smoothing. For example, suppose the underlying process undergoes an abrupt transition at time \(t_0\), moving from a regime with dense and large enough signals to a global null regime. In that case, tests based on local estimators may continue to reject some null hypotheses immediately after \(t_0\) because data from the pre-change region will leak into the post-change estimates. The \(\sup_t \mathrm{FDP}(t)\) may equal one even when false discoveries are well-controlled over nearly the entire interval. This distinction is illustrated in Figure~\ref{fig:supvsq}.

To address this issue, we introduce the asymptotically uniform false discovery rate (AuFDR), which can be viewed as a regularized version of the supremum FDP criterion. For any fixed $1\le r<\infty$, define $\|\mathrm{FDP}(\cdot)\|_{r}
=
\left\{\int_0^1 |\mathrm{FDP}(t)|^r\mathrm{d}t\right\}^{1/r}.$
For a pre-specified sequence $r_n\to\infty$, we define
\begin{align*}
\mathrm{AuFDR}_{r_n}
:=\mathbb{E}\left\{\|\mathrm{FDP}(\cdot)\|_{r_n}\right\}.
\end{align*}
Since $\|\mathrm{FDP}(\cdot)\|_{r_n}$ approaches $\sup_{t\in[0,1]}\mathrm{FDP}(t)$ as $r_n\to\infty$, the criterion retains a uniform interpretation. At the same time, the use of an $L_{r_n}$-norm avoids the excessive sensitivity to isolated local spikes. To illustrate this, suppose that outside a transition set $A_n\subset[0,1]$ with Lebesgue measure $\delta_n$, the FDP process is bounded by $\alpha$, whereas within $A_n$ it may be as large as one. Then
\begin{align*}
\|\mathrm{FDP}(\cdot)\|_{r_n}
\le
\alpha+\delta_n^{1/r_n}.
\end{align*}
Consequently, if $r_n=o\{\log(1/\delta_n)\}$, the influence of the transition set vanishes asymptotically. In contrast, $\sup_{t\in[0,1]}\mathrm{FDP}(t)$ remains uncontrolled whenever such a spike occurs. Thus, by employing a slowly diverging and user-specified value of $r_n$, the proposed $\mathrm{AuFDR}_{r_n}$ captures the global behavior of the FDP process while reducing sensitivity to local signal irregularities.

To illustrate the utility of this new framework, we develop an AuFDR control procedure for testing the null hypotheses of zero correlations over the rescaled time interval $t \in [0,1]$. The framework is designed for high-dimensional locally stationary time series and accommodates features common in neuroimaging \citep{lurie2020questions} and financial data \citep{jondeau2007financial, de2023modeling}, including mean discontinuities, non-Gaussianity, nonlinearity, and nonstationarity. 

To guarantee AuFDR control for this testing problem, we first construct time-varying $P$-values that are uniformly valid. The construction combines robust difference-based estimators with a multiplier bootstrap procedure. The difference-based estimators reduce sensitivity to unknown and potentially discontinuous mean functions, while the multiplier bootstrap approximates the maximum deviation of the relevant high-dimensional time-varying test statistics. Given these uniformly valid time-varying $P$-values, we develop a time-varying Benjamini--Yekutieli (BY) procedure for time-varying multiple testing. The procedure applies the BH step at the adjusted nominal level $\alpha/\ell(|\cH|)$, where $\ell(|\cH|)=\sum_{k=1}^{|\cH|}k^{-1}$, to account for arbitrary dependence. We show that the resulting procedure controls the $\mathrm{AuFDR}_{r_n}$ asymptotically with $r_n = o(\log n)$.

Our theoretical contribution is a new hyperrectangle Gaussian approximation result for high-dimensional \textit{second-order} structures of locally stationary processes. Existing Gaussian approximation results for locally stationary time series primarily focus on first-order statistics \citep{zhang2018gaussian,chang2024central}. Some recent work has begun to address second-order structures with maximum type Gaussian approximation results, see \cite{XiucaiDing2020} and \cite{zhang2024simultaneous} for examples. In contrast, correlation inference involves products of variables and therefore requires approximating the distribution of high-dimensional product processes over time. A related maximum-type approximation was developed by \citet{bai2025time} for uniform FWER control to obtain a single global threshold for the bootstrap method. However, in the present AuFDR framework, the time-varying BY procedure requires the simultaneous validity of all $P$-values, which naturally leads to Gaussian approximation over high-dimensional hyperrectangles. We establish such an approximation result and use it to prove the uniform validity of the proposed time-varying $P$-values, subsequently enabling us to establish AuFDR control for the time-varying BY procedure. \cite{zhang2024simultaneous}, \cite{XiucaiDing2020}

\subsection{Related work on  correlation testing and FDR control.}
Large-scale correlation testing has been extensively studied with time-invariant inference results \citep{efron2007correlation,cai2016large,bailey2019multiple}. These methods, however, are not designed for settings in which the correlation structure evolves over time. Recent work has addressed time-varying correlation inference with uniform family-wise error rate (FWER) control \citep{bai2025time}, but FDR-based control for time-varying correlation tests remains largely unexplored.

There is also a rich literature on FDR control under dependence \citep{benjamini2001control,barber2015controlling,fithian2022conditional}. Nevertheless, inference for time-varying correlations for high-dimensional and locally stationary time series  via multiple testing with FDR control poses several difficulties that are not covered by existing results. These include that (a) The underlying process is nonstationary, which leads to intricate limiting distributions of the test statistics that complicate testing the corresponding hypothesis. Moreover, the number of hypotheses diverges with dimension growing, which further sophisticates the joint distribution of the  $P$-values. (b) The time-varying $P$-values obtained in this setting exhibits complicated cross-sectional dependence across pairs together with intricate temporal dependence across the time index, which brings further difficulties on FDR control.
These features require a new analytical framework beyond existing results.

\begin{table}[h!]
\centering
\caption{Table of major notations for Section \ref{sec:DBmethod}.}
\label{tab:not1}
\begin{tabular}{l|l}
\hline
\textbf{Notation} & \textbf{Expression} \\
\hline
$Y_{j,i}, \epsilon_{j,i}, \mu_{i}(t_j)$ & the observation, innovation and trend function in \eqref{eq:model} \\
$h$ & Lag parameter in the difference-based method \\
$y_{j,i}$ & $Y_{j, i} - Y_{j-h, i}$ \\
$\hat \beta_{i,l}(t),\hat \beta'_{i,l}(t)$ & local linear estimators from \eqref{eq1} \\
$\tilde \gamma_{i,l}(t)$ & Estimator for pairwise covariance: $\hat \beta_{i,l}(t)/2$\\
$\tilde \sigma_{i,l}(t)$ & Estimator for marginal covariance: $\sqrt{\tilde \gamma_{i,i}(t)\tilde \gamma_{l,l}(t)}$ \\
$\tilde \rho_{i,l}(t)$ & Estimator for pairwise correlation: $\tilde\gamma_{i,l}(t)/\tilde\sigma_{i,l}(t)$ \\
\hline
\end{tabular}
\end{table}

\begin{table}[h!]
\centering
\caption{Table of major notations for Section \ref{sec:BootstrapInference} and \ref{sec:selectparam}.}
\label{tab:not2}
\begin{tabular}{l|l}
\hline
\textbf{Notation} & \textbf{Expression} \\
\hline
$e_{j,i,l}$ &  $(\epsilon_{j,i} - \epsilon_{j-h,i})(\epsilon_{j,l} - \epsilon_{j-h,l}) - \beta_{i,l}(t_j)$\\
$\hat{e}_{j,i,l}$ & Residuals of the regression in \eqref{eq1}: $y_{j,i}  y_{j,l} - \hat \beta_{i,l}(t_j)$\\
$\Xi_{j,i,l}$ & $e_{j,i,l}/2\sigma_{i,l}(t_j)- 4^{-1}\rho_{i,l}(t_j) \left\{e_{j, i,i}/\gamma_{i,i}(t_j) + e_{j,l,l}/ \gamma_{l,l}(t_j)\right\}$ \\
$\tilde\Xi_{j,i,l}$ & $y_{j,i}y_{j,l}/2\tilde\sigma_{i,l}(t_j)- 4^{-1}\tilde\rho_{i,l}(t_j) \left\{y_{j,i}^2/\tilde\gamma_{i,i}(t_j) + y_{j,l}^2/ \tilde\gamma_{l,l}(t_j)\right\}$ \\
$\hat \Xi_{j,i,l}$ & $2^{-1} \hat{e}_{j,i,l}/\tilde \sigma_{i,l}(t_j)-4^{-1}\tilde \rho_{i,l}(t_j) \big\{\hat{e}_{j,i,i}/\tilde \gamma_{i,i}(t_j) + \hat{e}_{j,l,l}/\tilde \gamma_{l,l}(t_j) \big\}$\\
$\vartheta_{i,l}(t)$ & $(nb_{i,l})^{-1}\sum_{j=1}^n K_{b_{i,l}}(t_j - t)\Xi_{j,i,l}$ \\
$\Gamma^2_{i,l}(t)$ & The limiting variance of $\sqrt{nb_{i,l}} \vartheta_{i,l}(t) $ \\
$\tilde \Delta_{s,i,l}$ & $\sum_{j=s}^{s+m_{i,l} -1} \hat{\Xi}_{j,i,l}$ \\
$\tilde{\Gamma}_{i,l}^{2}(t)$ & $\kappa m_{i,l}^{-1}\sum_{s=1}^{n}\Tilde \Delta^{2}_{s,i,l}\omega(t,s)$, ~$\kappa = \int K^2(t) \mathrm{d}t$,~ $\omega(t,s) = K_{\eta}(t-t_s)/\sum_{j=1}^n K_{\eta}(t_j -t)$ \\
$\bar {\Xi}_{j,s,i,l}$ & $c_{i,l} K_{b_{i,l}}(t_j - t_s)\Xi_{j,i,l}/\Gamma_{i,l}(t_s)$\\
$\hat{\bar {\Xi}}_{j,s, i,l}$ & $c_{i,l} K_{b_{i,l}}(t_j - t_s)\tilde{\Xi}_{j,i,l}/\tilde{\Gamma}_{i,l}(t_s)$ \\
$\hat{\bar{S}}_{j,s,i,l}$ & $\sum_{a = s-w+1}^{s} \hat{\bar{ \Xi}}_{a+j, \nb + j, i, l}-\sum_{a = s+1}^{s+w} \hat{\bar{\Xi}}_{a+j, \nb + j, i, l}$\\
$R_{s}^{(k)}$ & i.i.d. $N(0,1)$ random variables used in multiplier bootstrap \\
$Z_{\bt,i,l}^{(k)}$ & $\underset{j \in[ n- 2\lceil nb\rceil]}{\max}  \left| \sum_{s = w}^{ 2\lceil nb_{i,l}\rceil - w} \hat{\bar{S}}_{j, s, i, l}  R^{(k)}_{j+s}\right|/\sqrt{2 w\lceil nb\rceil }$\\
$T_{i,l}(t)$ & $(nb_{i,l})^{1/2}|\tilde \rho_{i,l}(t)|/\tilde{\Gamma}_{i,l}(t)$ \\
$P_{i,l}(t)$ & $B^{-1}\sum_{k=1}^B\mathbf{1}\sth{T_{i,l}(t)<Z_{\bt,i,l}^{(k)}}$\\
\hline
\end{tabular}
\end{table}

The rest of the paper is organized as follows. Section~\ref{sec:pre} formulates the testing problem and defines the AuFDR criterion. Section~\ref{sec:met} presents the proposed methodology, including the difference-based estimator, the multiplier bootstrap procedure, and the time-varying BY procedure. Section~\ref{sec:theo} establishes the theoretical guarantees, including the uniform validity of the time-varying $P$-values and asymptotic AuFDR control. Sections~\ref{Sec:Sim} and~\ref{sec:RDA} report simulation studies and real-data applications, respectively. Section \ref{data-availability-statement} contains the data availability statement. All technical proofs are delegated to the Appendix. In Table \ref{tab:not1} we list some notations used in time-varying correlation estimation in Section \ref{sec:DBmethod}. In Table \ref{tab:not2} we list the notations used in the inference procedure in Section \ref{sec:BootstrapInference} and \ref{sec:selectparam}.

\section{Preliminaries}\label{sec:pre}

\textbf{Notations.} For a positive integer $N$, let $[N] = \{1, 2, \dots, N\}$. For a vector $\mathbf{v} = (v_1, \dots, v_p)^\top \in \mathbb{R}^p$, let $|\mathbf{v}| = (\sum_{j=1}^p v^2_j)^{1/2}$, $|\mathbf{v}|_\infty = \max_{j} |v_j|$. For two vectors $\mathbf{v}, \mathbf{w} \in \mathbb{R}^p$, write $\mathbf{v} \le \mathbf{w}$ if $v_i \le w_i$ for all $i \in [p]$. For a random vector $\mathbf{V}$ and $q \geq 1$, let $\|\mathbf{V}\|_q = (\mathbb{E}|\mathbf{V}|^q)^{1/q}$. For a measurable function $f(t)$ on $t \in [0,1]$, we define its $L_r$-norm as $\|f\|_r = (\int_0^1 |f(t)|^r \mathrm{d}t)^{1/r}$. Throughout this paper, $K(\cdot)$ denotes a kernel function with support $(-1, 1)$, and we define $K_{b}(\cdot) = K(\cdot/b)$ for a bandwidth $b > 0$. For any set $B$, $|B|$ denotes its cardinality.

\subsection{High-dimensional Locally Stationary Time Series with Jumps in Mean}

We consider a $p$-dimensional time series $\{Y_{j,i}: i \in [p],\, j \in [n]\}$ with the following model
\begin{equation}\label{eq:model}
    Y_{j,i} = \mu_i(t_j) + \epsilon_{j,i}, 
    \qquad i \in [p],\; j \in [n],
\end{equation}
where $t_j = j/n$ is the rescaled observation time, $\mu_i:[0,1]\to \mathbb{R}$ is the deterministic mean function, and $\{\epsilon_{j,i}\}$ is the stochastic error process. The error process is assumed to be a non-Gaussian, nonlinear and locally stationary time series, and we consider a high-dimensional regime in which the dimension $p$ is allowed to grow with the sample size $n$. For mean functions, we suppose that $\mu_i$ has $d_i$ jumps:
\begin{equation}\label{eq:mean_piecewise}
    \mu_i(t)
    =
    \sum_{m=0}^{d_i}
    \mu_{i,m}(t)\,
    \mathbf{1}\bigl(a_{i,m} \le t < a_{i,m+1}\bigr),\notag
\end{equation}
where $0 = a_{i,0} < a_{i,1} < \cdots < a_{i,d_i} < a_{i,d_i+1} = 1$ are the jump locations for the $i$th component. On each $[a_{i,m}, a_{i,m+1})$, the function $\mu_{i,m}(\cdot)$ is assumed to be Lipschitz continuous, with Lipschitz constants uniformly bounded over $i \in [p]$ and $m \in [d_i]$. We allow $d_i$ to grow with $n$, corresponding to a regime of dense jumps.

Model \eqref{eq:model} also covers the special case of globally smooth mean functions with no structural breaks (i.e., $d_i = 0$). We focus on the more challenging setting where $d_i > 0$, while the case $d_i = 0$ can be treated similarly.

\subsection{Uniform Inference of Correlations and AuFDR}\label{sec:2.2}

For $i,l \in [p]$, define the time-varying covariance $\gamma_{i,l}(t)$, marginal variance $\sigma_{i,l}(t)$, and correlation $\rho_{i,l}(t)$ by
\begin{align*}
    \gamma_{i,l}(t) &= \lim_{n \rightarrow \infty} \mathrm{Cov}(\epsilon_{\lfloor nt \rfloor, i}, \epsilon_{\lfloor nt \rfloor, l}), \quad \sigma_{i,l}(t) = \sqrt{\gamma_{i,i}(t) \gamma_{l,l}(t)}, \quad \rho_{i,l}(t) = \gamma_{i,l}(t) / \sigma_{i,l}(t).
\end{align*}
We consider testing, for each $(i,l) \in \mathcal{H} = \{(i,l): 1 \le l < i \le p\}$ and each $t \in [0,1]$, the hypotheses
\begin{equation}\label{eq:hypotheses}
    H_{i,l,t}^0: \rho_{i,l}(t) = 0
    \quad \text{vs.} \quad
    H_{i,l,t}^1: \rho_{i,l}(t) \ne 0.
\end{equation}
Let $\mathcal{R}(t) \subset \mathcal{H}$ denote the set of rejected hypotheses at time $t$, and let $\mathcal{H}_0(t) = \{(i,l) \in \mathcal{H} : \rho_{i,l}(t) = 0\}$
denote the set of true null hypotheses. The FDP at time $t$ is defined by
\begin{equation}\label{eq:defFDP}
    \mathrm{FDP}(t) = \frac{|\mathcal{H}_0(t) \cap \mathcal{R}(t)|}{\max\{|\mathcal{R}(t)|, 1\}}.
\end{equation}
To provide simultaneous error control over the entire FDP trajectory, we define the AuFDR as
\begin{equation}\label{eq:defFDR}
    \mathrm{AuFDR}_{r_n} = \mathbb{E}\left\{ \| \mathrm{FDP}(\cdot) \|_{r_n} \right\},
\end{equation}
where $r_n\rightarrow\infty$ is a sequence of parameters chosen by the user. Our goal is to develop a procedure that controls $\mathrm{AuFDR}_{r_n}$ at a prescribed level.

\begin{remark}[Network visualization]
    The proposed framework naturally yields a time-varying correlation network $\mathcal{G}(t) = \{\mathcal{V}, \mathcal{E}(t)\}$, where $\mathcal{V} = [p]$ is the vertex set, and the time-varying edge set $\mathcal{E}(t) = \{(i,l) \in \mathcal{H} : H_{i,l,t}^0 \text{ is rejected}\}$ represents pairs with statistically significant nonzero correlation. 
\end{remark}

\begin{remark}[Other hypothesis testing problems]
    While we focus on testing the null hypothesis of zero correlation as defined in \eqref{eq:hypotheses}, the proposed framework can be readily extended to other hypotheses. For instance, one could test $H_{i,l,t}^{0\prime}: \rho_{i,l}(t) = g_{i,l}(t)$, where $g_{i,l}(t)$ is any given measurable function.
\end{remark}

\begin{remark}[Difficulty of controlling AuFDR]\label{remark:3}
    Controlling $\mathrm{AuFDR}_{r_n}$ is more challenging than obtaining pointwise FDR control over local windows, because pointwise control of the FDP does not imply control of the $L_{r_n}$-norm of the time-varying FDP when $r_n\rightarrow\infty$. By comparison, our inference procedure can  control $\mathrm{AuFDR}_{r_n}$.
\end{remark}

\section{Methodology}\label{sec:met}

\subsection{Difference-based Correlation Estimation Method}\label{sec:DBmethod}

To estimate the time-varying correlation function $\rho_{i,l}(t)$ in the presence of abrupt jumps in means, we adopt a difference-based approach in \cite{bai2025time}. Let $h = h_n$ be a sequence of integers such that $h \to \infty$ and $h = o(n)$. For $j\in\{h+1,\dots,n\}$ and $i,l\in[p]$, define $\beta_{i,l}(t_{j}) := \eE\sth{(\epsilon_{j,i} - \epsilon_{j-h,i})(\epsilon_{j,l} - \epsilon_{j-h,l})}$. Under the assumption that the temporal dependence of the error process decays fast enough as the lag $h$ increases, we have $\mathbb{E}(\epsilon_{j,i}\epsilon_{j-h,l}) \approx 0$, which implies $\beta_{i,l}(t_{j}) \approx 2\gamma_{i,l}(t_j)$. Furthermore, provided the trend function $\mu_i(\cdot)$ is piecewise Lipschitz and the lag $h$ is sufficiently small relative to $n$, the differenced observations $y_{j,i} := Y_{j,i} - Y_{j-h,i}$ effectively filter out the mean trend: $y_{j,i} \approx \epsilon_{j,i} - \epsilon_{j-h,i}$, except in the vicinity of a jump.

Based on the analysis above, we estimate $\gamma_{i,l}(t)$ by the local linear estimator $\tilde \gamma_{i,l}(t)  = \hat \beta_{i,l}(t)/2$, with \begin{align}\label{eq1}
    &\sth{\hat \beta_{i,l}(t),\hat \beta_{i,l}^{\prime}(t)}^{\T}  = \underset{(\eta_0, \eta_1) \in \eR^2}{\mathrm{argmin}}\sum_{j=h+1}^n\sth{y_{j, i} y_{j,l} - \eta_0 - \eta_1(t_j - t)}^2 K_{b_{i,l}}(t_j - t),
\end{align}
where $b_{i,l}$ is the bandwidth parameter.  To simplify notation, we suppress the dependence of the estimator on $h$. The choice of $h$ is discussed in Section \ref{sec.3.4}.
 The time-varying correlation function $\rho_{i,l}(t)$ is subsequently estimated by
\begin{align}\label{eq2}
      \tilde \rho_{i,l}(t) = \tilde \gamma_{i,l}(t)/\tilde \sigma_{i,l}(t),~~\mathrm{with}~~\tilde \sigma_{i,l}(t) = \sqrt{\tilde \gamma_{i,i}(t) \tilde \gamma_{l,l}(t)}.
\end{align}

\begin{remark}[Comparison with jump detection]
    A conventional alternative for estimating time-varying correlation functions with possible jumps is a three-step procedure: (i) detect all jumps, (ii) estimate the mean function separately within each resulting segment, and (iii) remove the estimated mean before and then conduct inference for time-varying correlations. The segmentation step can create short sub-intervals and exacerbate boundary effects, which can substantially degrade the performance of local smoothing methods and subsequent time-varying correlation estimation. By contrast, difference-based methods avoid explicit jump estimation and have proved effective in estimating second order structures under various scenarios of trends, especially when the trends are time-varying and discontinuous. \citep{muller1987estimation, hall1990asymptotically, tecuapetla2017autocovariance, dette2019detecting, cui2021estimation, Chan2022}.
\end{remark}

\subsection{Generate Uniformly Valid $P$-values via Bootstrap}\label{sec:BootstrapInference}

To achieve AuFDR control for hypothesis testing in \eqref{eq:hypotheses}, we need to go beyond pointwise inference and conduct simultaneous inference, as noted in Remark \ref{remark:3}. This requires us to characterize the distribution of \begin{align}\label{eq:maxdev}
    \sup_{t\in[b_{i,l},1-b_{i,l}]} (nb_{i,l})^{1/2}|\tilde{\rho}_{i,l}(t) - \rho_{i,l}(t)|/\Gamma_{i,l}(t),
\end{align}
where $\Gamma_{i,l}(t)$ is the asymptotic standard deviation of $\tilde{\rho}_{i,l}(t)$. Since the distribution of this supremum statistic is not available in closed form, we employ a multiplier bootstrap to approximate it. The resulting bootstrap-based $P$-values are asymptotically uniformly valid and can therefore be used in the subsequent AuFDR control procedure.

Our construction relies on a stochastic expansion for $\tilde\rho_{i,l}(t)$, adopted from the framework in \citet{bai2025time}. Define $e_{j,i,l} = (\epsilon_{j,i} - \epsilon_{j-h,i})(\epsilon_{j,l} - \epsilon_{j-h,l}) - \beta_{i,l}(t_j)$, and
\begin{align}
    \vartheta_{i,l}(t) &:= (nb_{i,l})^{-1}\sum_{j=1}^n  K_{b_{i,l}}(t_j - t)\Xi_{j,i,l},\label{eq8}\\
   \Xi_{j,i,l} &:= 2^{-1} e_{j,i,l}/\sigma_{i,l}(t_j) - 4^{-1}\rho_{i,l}(t_j) \sth{e_{j,i,i}/\gamma_{i,i}(t_j) + e_{j,l,l}/ \gamma_{l,l}(t_j)},\label{eq:Xidiscrete}
 \end{align}
 where $\{\Xi_{j,i,l}\}_{j=1}^n$ is a locally stationary innovation process. Following \eqref{pfthm1eq2} and \eqref{pfthm1eq3} of the Appendix,  we have \begin{align}\label{eq:approx3}
     \max_{i,l\in[p]}&\sup_{t \in [b_{i,l},1-b_{i,l}]} (nb_{i,l})^{1/2}\left|\tilde \rho_{i,l}(t)-\rho_{i,l}(t)-\vartheta_{i,l}(t)\right| = o_{\eP}(1).
 \end{align} Let $b = \max_{(i,l)\in\mathcal{H}} b_{i,l}$ and $c_{i,l} = (b/b_{i,l})^{1/2}$. We define the long run variance of $\vartheta_{i,l}(t)$ and the standardized version of $\{\Xi_{j,i,l}\}_{j=1}^n$ respectively as:
 \begin{align}
\Gamma_{i,l}^2(t)&:=\lim_{n\rightarrow\infty}\var\sth{(nb_{i,l})^{1/2}\vartheta_{i,l}(t)}, ~~\bar {\Xi}_{j,s,i,l} := c_{i,l} K_{b_{i,l}}(t_j - t_s)\Xi_{j,i,l}/\Gamma_{i,l}(t_s).\label{eq:defXjsz}
\end{align}
By \eqref{eq8}, \eqref{eq:Xidiscrete}, \eqref{eq:approx3} and \eqref{eq:defXjsz}, we have the following approximation intuitively with negligible error: \begin{align*}
    \sup_{t \in [b_{i,l},1-b_{i,l}]} (nb_{i,l})^{1/2}\left|\tilde \rho_{i,l}(t)-\rho_{i,l}(t)\right|/\Gamma_{i,l}(t)\approx \underset{1 \leq j\leq n}{\max}  \left|\sum_{s=1}^{n}\bar{ \Xi}_{j,s,i,l}/\sqrt{nb}\right|_{\infty}.
\end{align*}

To test $H^0_{i,l,t}$, we use the test statistic
\begin{align}
T_{i,l}(t)
=(nb_{i,l})^{1/2}|\tilde\rho_{i,l}(t)|/\tilde\Gamma_{i,l}(t),
\label{eq:test-stat}
\end{align}
where $\tilde\Gamma_{i,l}(t)$ is a consistent estimator of $\Gamma_{i,l}(t)$. Under the null hypothesis $H^0_{i,l,t}:\rho_{i,l}(t)=0$, the statistic in \eqref{eq:test-stat} coincides with the standardized estimation error, i.e. \eqref{eq:maxdev}, up to negligible terms. 

\begin{algorithm}[htbp]
    \caption{Inference of time-varying correlations with AuFDR control}
    \begin{algorithmic} [1]
      \State \textbf{Input:} Observed data $Y_{j,i},j\in[n],i\in[p]$.     
      \State Compute $\tilde \rho_{i,l}(t)$ for $(i,l)\in\cH$ as defined in \eqref{eq2}, and compute $\hat{\bar{ \Xi}}_{j,s,i,l}$ in \eqref{eq:hatxijsz} for $j,s\in [n]$ and $(i,l)\in\cH$.
        \State For a window size $w$, compute $\hat{\bar{S}}_{j,s,i,l}= \sum_{a = s-w+1}^{s} \hat{\bar{ \Xi}}_{a+j, \nb + j, i, l}-\sum_{a = s+1}^{s+w} \hat{\bar{\Xi}}_{a+j, \nb + j, i, l}$ for $j\in[n - 2\nb]$, $s\in[w,2\nb-w]$, and $(i,l)\in\cH$.
        \For{$k = 1, \cdots, B$}
        \State Generate independent standard normal random variables $R_s^{(k)}$, $s\in[n]$. 
        \State Define $b = \max_{(i,l)\in\cH}b_{i,l}$. For each $(i,l)\in\cH$, calculate
          $$Z_{\bt,i,l}^{(k)} =\frac{\underset{j \in[ n- 2\lceil nb\rceil]}{\max}  \left| \sum_{s = w}^{ 2\lceil nb_{i,l}\rceil - w} \hat{\bar{S}}_{j, s, i, l}  R^{(k)}_{j+s}\right|}{\sqrt{2 w\lceil nb\rceil }}.$$
    \EndFor
    \State Calculate test statistics \begin{align}
        T_{i,l}(t) = (nb_{i,l})^{1/2}|\tilde \rho_{i,l}(t)|/\tilde{\Gamma}_{i,l}(t)\label{eq:Alg1eq1}
    \end{align}
    and compute the time-varying $P$-values for testing $H_{i,l,t}^0$ as \begin{align*}
        P_{i,l}(t) = \frac{1}{B}\sum_{k=1}^B\mathbf{1}\sth{T_{i,l}(t)<Z_{\bt,i,l}^{(k)}}.
    \end{align*}
    \State For each time point $t$, let $P_{(1)}(t) \le \cdots \le P_{(|\cH|)}(t)$ denote the pointwise order statistics of the $P$-values. For a target AuFDR level $\alpha$, reject the hypotheses with $P$-values not exceeding $R(t)\alpha/{|\cH|\ell(|\cH|)}$, where $\ell(|\cH|)=\sum_{j=1}^{|\cH|}j^{-1}$ and $R(t) := \max\sth{k:P_{(k)}(t) \le \alpha k/{|\cH|\ell(|\cH|)}}$, with $R(t)=0$ if the set is empty.
    \State  \textbf{Output:} Test results $\cR(t) = \sth{(i,l)\in\cH: H_{i,l,t}^0 ~\mbox{is rejected}}$ at all time $t$.
    \end{algorithmic}
    \label{alg1}
\end{algorithm}

Algorithm~\ref{alg1} summarizes the proposed inference procedure. In Steps 4--7, we generate bootstrap replicates $
Z_{\mathrm{boot},i,l}^{(1)},\dots,Z_{\mathrm{boot},i,l}^{(B)},
$ which approximate the null distribution of $\sup_{t\in[b,1-b]} T_{i,l}(t)$. For each $(i,l)\in\mathcal H$ and $t\in[0,1]$, we then define the bootstrap $P$-value as
\begin{align}
P_{i,l}(t)
=
\frac{1}{B}\sum_{k=1}^B
\mathbf 1\left\{
T_{i,l}(t)< Z_{\mathrm{boot},i,l}^{(k)}
\right\}.
\label{eq:defpval}
\end{align}
Thus, for each $(i,l)\in\cH$, \(P_{i,l}(t)\) uniformly consistently estimates the upper-tail probability of the observed
statistic under the null distribution of maximum deviation over $t\in[b,1-b]$, with small values providing
evidence against \(H^0_{i,l,t}\). These $P$-values are subsequently combined through a time-varying BY procedure to achieve AuFDR control under arbitrary dependence.

\begin{figure}[htbp]
    \centering
    \includegraphics[width=0.9\linewidth]{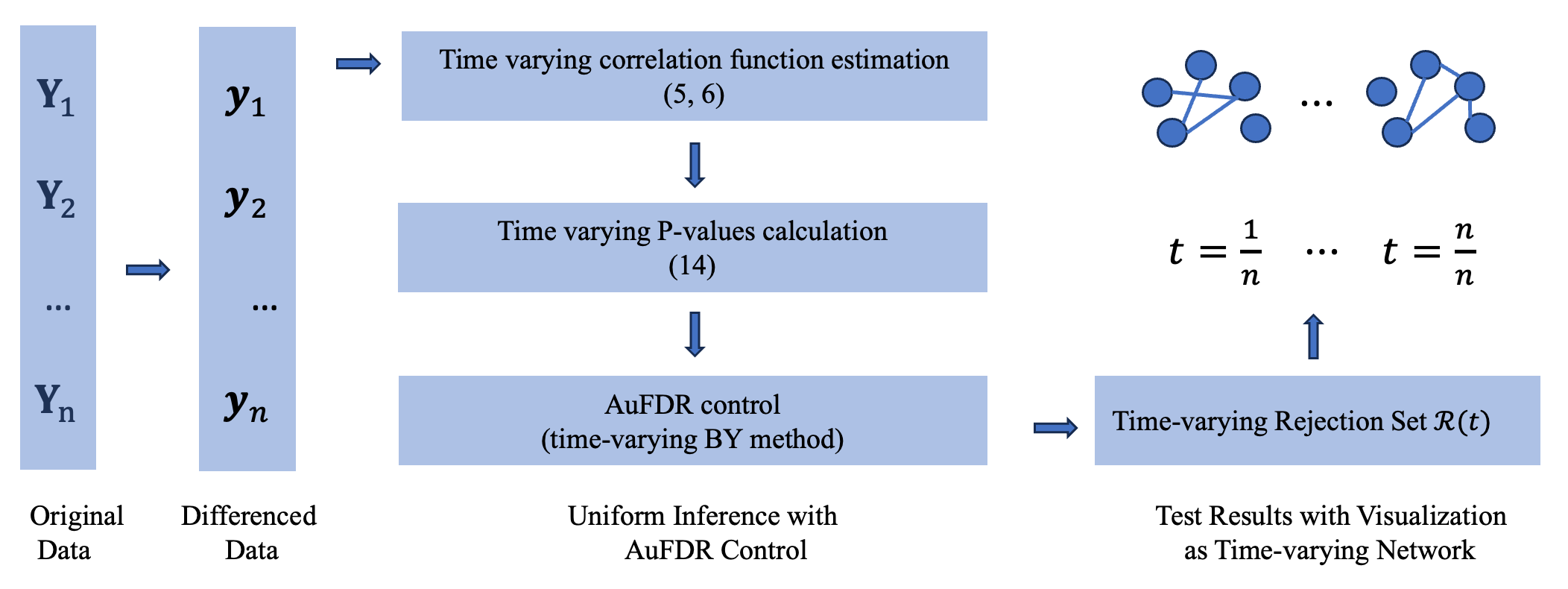}
    \caption{A schematic diagram of Algorithm \ref{alg1}.}
\end{figure}

\subsection{Estimation of quantities in the bootstrap procedure}\label{sec:selectparam}

To implement the multiplier bootstrap, we estimate the innovation process $\Xi_{j,i,l}$, the variance function $\Gamma_{i,l}^2(t)$, and the standardized local innovations $\bar\Xi_{j,s,i,l}$ in \eqref{eq:Xidiscrete} using residual-based estimators, following \citet{bai2025time}. For estimation of $\Xi_{j,i,l}$, we use the following difference-based estimator:
\begin{align}
    &\tilde  \Xi_{j,i,l} = \frac{y_{j,i} y_{j,l}}{2\tilde\sigma_{i,l}(t_j)} -\frac{\tilde \rho_{i,l}(t_j)}{4} \sth{ y_{j,i}^2/\tilde \gamma_{i,i}(t_j) + y_{j,l}^2/\tilde \gamma_{l,l}(t_j)},\notag
 \end{align}
where $y_{j,i} = Y_{j,i} - Y_{j-h, i}$.
Next, we estimate the limiting variance function $\Gamma_{i,l}^2(\cdot)$. Let $\hat{e}_{j,i,l} = y_{j,i}  y_{j,l} - \hat \beta_{i,l}(t_j)$ denote the residuals from the regression in \eqref{eq1}. Let $m_{i,l}$ be the smoothing-window length, and let $\eta$ be a bandwidth parameter. We estimate $\Gamma_{i,l}^2(\cdot)$ by
\begin{align}\label{eq7}
    \tilde{\Gamma}_{i,l}^{2}(t) &=  \frac{\kappa}{m_{i,l}}\sum_{s=1}^{n}\Tilde \Delta^{2}_{s,i,l}\omega(t,s),
\end{align}
where $\omega(t,s) = K_{\eta}(t-t_s)/\sum_{j=1}^n K_{\eta}(t_j -t)$, $\kappa = \int K^2(t)\mathrm{d}t$  , $\Tilde \Delta_{s,i,l} = \sum_{j=s}^{s+m_{i,l} -1} \hat \Xi_{j,i,l}$, and $\hat \Xi_{j,i,l} = 2^{-1} \hat{e}_{j,i,l}/\tilde \sigma_{i,l}(t_j)-4^{-1}\tilde \rho_{i,l}(t_j) \big\{\hat{e}_{j,i,i}/\tilde \gamma_{i,i}(t_j) + \hat{e}_{j,l,l}/\tilde \gamma_{l,l}(t_j) \big\}$. Finally, for $j,s\in[n]$, we estimate $\bar {\Xi}_{j,s,i,l}$ by
\begin{align}\label{eq:hatxijsz}
   \hat{\bar {\Xi}}_{j,s, i,l} &:= c_{i,l} K_{b_{i,l}}(t_j - t_s)\tilde{\Xi}_{j,i,l}/\tilde{\Gamma}_{i,l}(t_s).
\end{align}

\subsection{AuFDR Control via Time-varying BY Procedure}

To control the AuFDR under arbitrary dependence, we apply a time-varying BY procedure \citep{benjamini2001control} to the time-varying $P$-values. Let $P_{(1)}(t) \le P_{(2)}(t) \le \dots \le P_{(|\cH|)}(t)$ denote the ordered $P$-values at a fixed time point $t \in [0,1]$, and let $\ell(|\cH|)=\sum_{k=1}^{|\cH|}k^{-1}\asymp\log(|\cH|)$. For target level $\alpha$, the BY procedure at time $t$ determines the rejection threshold by identifying the largest integer $R(t)$ such that
\begin{equation*}
    R(t) = \max \left\{ k : P_{(k)}(t) \le \frac{\alpha k}{|\cH|\ell(|\cH|)} \right\}.
\end{equation*}
If no such $k$ exists, we set $R(t) = 0$. We then reject all null hypotheses $H^0_{i,l,t}$ whose corresponding $P$-values satisfy $P_{i,l} (t) \le \alpha R(t) /\{|\cH|\ell(|\cH|)\}$.

A key feature of our framework is that pointwise application of the BY procedure is sufficient for AuFDR control. This is because the $P$-values, as defined in \eqref{eq:defpval}, are constructed from $Z^{(1)}_{\bt,i,l},\dots,Z^{(B)}_{\bt,i,l}$, which approximate the distribution of the maximum deviation. This construction yields \textit{simultaneous} validity of the time-varying $P$-values over the time interval of $t$ where $H_{i,l,t}^0$ is true, see Theorem \ref{thm:pvalp}.

\begin{comment}[Connection between Benjamini--Hochberg (BH) and BY procedure]
Similar to the BY procedure, the Benjamini--Hochberg (BH) procedure is also a classical  FDR control procedure: after ordering the $P$-values, it determines the rejection threshold by choosing an integer $R(t)$ such that \begin{align*}
    R(t) = \max\left\{k:P_{(k)}(t)\le \alpha k/|\cH|\right\},
\end{align*}
and rejects all null hypotheses $H_{i,l,t}^0$ whose $P$-values $P_{i,l}(t) \le \alpha R(t)/|\cH|$. It controls the FDR under independence or positive regression dependence on a subset (PRDS) \citep{benjamini2001control}. In our setting, these conditions are often violated because the time-varying $P$-values share common bootstrap samples and inherit complex cross-sectional dependence across $(i,l)\in\cH$ and temporal dependence across $t\in[0,1]$. We therefore use the BY procedure, which lowers the nominal level in the BH procedure by the harmonic factor $\ell(|\cH|)$. This adjustment allows us to work without dependence restrictions and yields the asymptotic AuFDR bound in Theorem~\ref{thm3}.
\end{comment}

\subsection{Tuning Parameters Selection}\label{sec.3.4}

When implementing Algorithm \ref{alg1}, we need to select the following parameters: the lag $h$, the bandwidths $b_{i,l}$ used to estimate $\rho_{i,l}(t)$, the window length $m_{i,l}$ and bandwidth $\eta$ used to estimate $\Gamma_{i,l}(t)$, and the window length $w$ used in Step 3 of Algorithm \ref{alg1}.

For the choice of lag $h$, the sensitivity analysis in Section \ref{sec:Sensi} shows that Algorithm \ref{alg1} is robust to a range of choices of $h$. Assumption \ref{Ass:bdrate} requires $h\asymp \log n$, and hence in practice one may use $h = \lceil a\log n \rceil$ for a manually set $a>0$. In the simulation studies reported in this paper, we use $h = \lceil \log n \rceil$.

For the selection of $b_{i,l},i,l\in[p]$, we employ the generalized cross validation (GCV) method proposed by \cite{wahba1975smoothing}. We choose $b_{i,l}$ by minimizing \begin{align*}
    \mathrm{GCV}(b) = \frac{n^{-1}|\bY^\diamond_{i,l}-\widehat{\bY}^\diamond_{i,l}|^2}{\qth{1-\mathrm{tr}\sth{\bQ_{i,l}(b)}/n}^2},
\end{align*}
where $\bQ_{i,l}(b)$ is a square matrix that depends on $b$ satisfying $\widehat{\bY}^\diamond_{i,l} = \bQ_{i,l}(b)\bY_{i,l}^\diamond$. Recall that $y_{j,i} = Y_{j,i}-Y_{j-h,i}$. We take
$\bY_{i,l}^\diamond = (y_{1,i}y_{1,l},\dots,y_{n,i}y_{n,l})^\T$
and
$\widehat{\bY}^\diamond_{i,l} = (\hat{\beta}_{i,l}(t_1),\dots,\hat{\beta}_{i,l}(t_n))^\T$.

To select $\eta$ and $w$, we use the extended minimum volatility (MV) method in Chapter 9 of \cite{politis1999subsampling}. We first propose a grid of candidate window sizes $w_1,\dots,w_{M_1}$ and bandwidths $\eta_1,\dots,\eta_{M_2}$. In this step, we first fix $m = m_0 = \lfloor n^{2/7}\rfloor$. Denote the sample variance of the bootstrap statistics given the dataset by $s^2_{w_{j_1},\eta_{j_2}}$, i.e., \begin{align*}
    s^2_{w_{j_1},\eta_{j_2}} = \sum_{s=w_{j_1}}^{n-w_{j_1}}\sum_{j=0}^{n-2\nb} \sum_{(i,l)\in \cH} \hat{\tilde{S}}_{j,s,i,l}^2(w_{j_1},\eta_{j_2},m_0),\quad j_1\in[M_1],~j_2\in[M_2],
\end{align*}
where $\hat{\tilde{S}}_{j,s,i,l}(w_{j_1},\eta_{j_2},m_0)$ is the estimator in Step 3 of Algorithm \ref{alg1} using $w_{j_1}$ in the block sum, and $m_0,\eta_{j_2}$ in the estimator $\tilde{\Gamma}_{i,l}(t)$ in \eqref{eq7}. We then choose the pair $(j_1,j_2)$ that minimizes the local variability criterion
\begin{align*}
    \mathrm{MV}(j,j') := \mathrm{SD}\sth{\pth{\cup_{r=-1}^{1}s^2_{w_j,\eta_{j'}+r}}\cup \pth{\cup_{r=-1}^{1}s^2_{w_{j+r},\eta_{j'}}}},
\end{align*}
where $\mathrm{SD}$ denotes the sample standard deviation. Given the selected $w_{j_1},\eta_{j_2}$, we could further refine $m_{i,l}$ using an additional MV step. Specifically, we consider a grid of candidate block sizes
$m_{i,l,1},\ldots,m_{i,l,M_3}$ and select the index $j_3$ that minimizes
\begin{align*}
    \mathrm{SD}(j)
    =
    \int_0^1
    \mathrm{SD}\sth{
    \tilde{\Gamma}_{i,l}^2(t;m_{i,l,j+r},\eta_{j_2}),
    ~r=-1,0,1
    }
    \mathrm{d}t.
\end{align*}
The final choice is then $m_{i,l}=m_{i,l,j_3}$.

\section{Theoretical Results}\label{sec:theo}

We assume that for each dimension $i \in [p]$, the error process $(\epsilon_{j,i})_{j \in [n]}$ admits the representation:
\begin{equation}
    \epsilon_{j,i} = G_i(t_j, \cF_j),
\end{equation}
where $t_j = j/n$, $G_i(\cdot, \cdot)$ are measurable functions $[0,1] \times \eR^{\eZ} \rightarrow\eR$, and $\cF_j = (\dots, \xi_{j-1}, \xi_j)$ are the filtration of i.i.d. random elements $(\xi_j)_{j \in \eZ}$. This setup is flexible and encompasses time-varying AR, ARCH, and GARCH models \citep{Dannazhang2021}.

For a process $G(t, \cF_j)$, we say that it is $\cL^q$ stochastic Lipschitz continuous (denoted as $G \in \mathrm{Lip}_q$) if there exists a constant $C > 0$ such that for any $t_1, t_2 \in [0,1]$, $\|G(t_1, \cF_0) - G(t_2, \cF_0)\|_q \le C |t_1 - t_2|$. For $q \ge 2$, such processes are named locally stationary processes. To quantify temporal dependence, let $\cF_j^* = (\dots, \xi_{-1}, \xi_0', \xi_1, \dots, \xi_j)$ be a coupled version of $\cF_j$, where $\xi_0$ is replaced by an i.i.d. copy $\xi_0'$. The physical dependence measure \citep{wu2005nonlinear} is defined by
\begin{equation}
    \delta_q(G, l) = \sup_{t \in [0,1]} \|G(t, \cF_l) - G(t, \cF_l^*)\|_q, \quad l \ge 0,
\end{equation}
and $\delta_q(G, l) = 0$ for $l < 0$.

We propose the following assumptions about the error processes, the trend functions with jumps, the kernel and bandwidths.

\begin{assumption}  \label{Ass:error}

    (\romannumeral 1) For some $q>4$, $G_i(\cdot,\cdot) \in \mathrm{Lip}_{2q}$ with uniformly bounded Lipschitz constants.

    (\romannumeral 2) There exists $t_0 > 0, \kappa\in(0,1)$ such that $\eE\sth{\exp(t_0|G_i(t, \cF_0)|^{2\kappa})}$ is uniformly upper bounded for all $t\in[0,1],i\in[p]$.

    (\romannumeral 3) For some $\chi \in (0,1), q>4$, $\max_{i\in[p]}\delta_{2q}(G_i, r) = O(\chi^r)$.
    
    (\romannumeral 4) For all $i,l\in[p]$, $\gamma_{i,l}(\cdot)$ and their second derivative $\gamma_{i,l}^{\prime\prime}(\cdot)$ are Lipschitz continuous on $[0,1]$ with uniformly upper bounded Lipschitz constants, and $\gamma_{i,i}(\cdot)$ are uniformly bounded away from $0$.

    (\romannumeral 5)  For all $i,l\in[p], t\in[0,1]$, $\Gamma^{2}_{i,l}(t)$ in \eqref{eq:defXjsz} are well defined and uniformly upper bounded and lower bounded away from $0$. Their derivatives $\Gamma_{i,l}^{\prime}(t)$ are uniformly upper bounded.

    (\romannumeral 6) The maximum number of jumps $d:=\max_{1 \leq i \leq p} d_i = O(n^{\phi})$, $0 \leq \phi < 1/5$, and for an absolute constant $C>0$, $\underset{1\leq i  \leq p, 1\leq m \leq d_i}{\max} |\mu_{i}(a_{i,m}^{+}) -\mu_{i}(a_{i,m}^{-})| \leq C$.

    (\romannumeral 7) The kernel $K(\cdot)$ is symmetric with support $[-1,1]$, and its second order derivative $K^{\prime\prime}$ is Lipschitz continuous on $[-1,1]$. Furthermore, $\int_{-\infty}^{\infty} K(u) du = 1$, and $\int_{-\infty}^{\infty} (K^{\prime}(u))^2 du < \infty$. 

    (\romannumeral 8) There exists a constant $C >0$ such that $\min_{i,l\in[p]}b_{i,l}/b\geq C$ with $b = \max_{i,l\in[p]} b_{i,l}$. 
\end{assumption}

    Assumptions \ref{Ass:error}(\romannumeral 1) and \ref{Ass:error}(\romannumeral 3) are standard in the kernel-based nonparametric analysis of locally stationary time series, see Assumption 1 of \cite{zhao2015inference} for example. Assumption \ref{Ass:error}(\romannumeral 2) allows sub-Weibull-type tails (which could be weaker than sub-exponential conditions). Assumption \ref{Ass:error}(\romannumeral 4) and \ref{Ass:error}(\romannumeral 5) guarantee that $\gamma_{i,l}(t), \Gamma_{i,l}(t)$'s vary smoothly with respect to $t$.  Assumption \ref{Ass:error}(\romannumeral 6) accommodates scenarios with an increasing number of jumps. Assumptions \ref{Ass:error}(\romannumeral 7) and \ref{Ass:error}(\romannumeral 8) are standard regularity conditions on the kernel and the bandwidths.

    Recall that $w$ is the block size in Step 3 of Algorithm \ref{alg1}, $m_{i,l}$ and $\eta$ are the smoothing parameters for the estimator of $\Gamma_{i,l}(\cdot)$, see \eqref{eq7}. 
    Let $m = \max_{i,l\in[p]} m_{i,l}$, $\vartheta_n = \frac{\log^2 n}{w} + \frac{w}{nb} + (\frac{w}{nb})^{1/2}(np^2)^{4/q}$, $g_n = p^{2/q}\big\{(\frac{m}{n \eta^{2}})^{1/2}+m^{-1}+\eta  +(\frac{m}{nb})^{1/2}(\frac{mb}{n})^{-1/(2q)}+ (nb)^{-1/2}b^{-1/q}\big\} + w^{3/2}/n + n^{\phi}h/\sqrt{w}$.

    \begin{assumption}\label{Ass:bdrate}
    The bandwidth satisfies $b \asymp n^{c}$ for some constant $c$ satisfying $-1/4 < c \le -1/5$, and the lag satisfies $h\asymp \log n$. Moreover, define $c_n = (n^{\phi-1}b^{-1}h + b^{2})(p^2/b)^{1/q}$, and  \begin{align}
        &p^2/(n^{q/2+\phi}b^{q/2+1}h+b^{q/2+4}n^{1+q/2})\rightarrow 0,\notag\\
        &\theta_n^{(1)} := (nb)^{-\frac{1}{8}}\pth{\log n}^{4} + \sth{(\sqrt{nb}c_n)^{\frac{q}{q+1}}+(p^2/(nb)^q)^{\frac{1}{q+1}}}\pth{\log n}^{1/2}\rightarrow 0,\notag\\
   &\theta_n^{(2)}:=\vartheta_n^{1/3}\pth{\log n}^{2/3}+\big(g_n (np^4)^{1/q}\big)^{q/(q+2)}\log n \to 0.\label{bdrateeq}
    \end{align}
    \end{assumption} 

Assumption \ref{Ass:bdrate} can be satisfied for sufficiently large $q$ if $w\asymp n^{2/5},\eta\asymp n^{-1/7}, b\asymp n^{-1/5}, m\asymp n^{2/7}, \phi < 1/5$ and $p\asymp n^{\iota'}$, where $\iota' > 0$ is any fixed positive number. It allows $p$ to increase polynomially with respect to the sample size $n$.

    Recall the definition of $\bar { \Xi}_{j,s, i,l}$ in \eqref{eq:defXjsz}. For any $(i,l)\in\cH$, define an $(n-2\nb + 1)$-dimensional vector $\bar{\mf \Xi}_{j,i,l} = (\bar{ \Xi}_{j,\nb,i,l}, \bar{\Xi}_{j + 1,\nb + 1, i, l}, \cdots, \bar{\Xi}_{n - 2\nb +j, n - \nb,i,l})^{\T}$, and an $(n-2\nb + 1)|\cH|$-dimensional vector $\bar{\mf \Xi}_{j}^{\cG} = (\bar{\mf \Xi}_{j,2,1},\bar{\mf \Xi}_{j,3,1},\bar{\mf \Xi}_{j,3,2},\dots,\bar{\mf \Xi}_{j,p,p-1})^{\T}$. To simplify notation, for any arbitrary $L\in\eN$ and $L|\cH|$-dimensional vector $\bZ = (\bZ_{2,1},\bZ_{3,1},\bZ_{3,2},\dots,\bZ_{p,p-1})^\T$ with $\bZ_{i,l} = (Z_{1,i,l},\dots, Z_{L,i,l})^\T$, define \begin{align}\label{eq6}
    \ath{\bZ}_{\infty,p} = \Big(\max_{j\in[L]}|Z_{j,2,1}|,\max_{j\in[L]}|Z_{j,3,1}|,\max_{j\in[L]}|Z_{j,3,2}|,\dots,\max_{j\in[L]}|Z_{j,p,p-1}|\Big)^\T.
\end{align}

\begin{theorem}[Gaussian Approximation]\label{thm1}
    Assume that Assumption \ref{Ass:error} and \ref{Ass:bdrate} hold, and there exist constants $\iota,\iota' > 0$ such that $p = O(n^\iota)$ and $b\asymp n^{-\iota'}$. Let $\bx = (x_{2,1},x_{3,1},x_{3,2},\dots,x_{p,p-1})\in \eR^{|\cH|}$, and $\cT = [b,1-b]$. Then there exists a sequence of zero-mean Gaussian vectors $(\mf Z_j)_{j=1}^{2\nb} = (\mf Z_{j,2,1},\dots,\mf Z_{j,p,p-1})_{j=1}^{2\nb} \in \eR^{(n - 2\nb+1)|\cH|}$, which share the same autocovariance structure with the vectors $(\bar{\mf \Xi}^{\cG}_j)_{j=1}^{2\nb} $, such that
\begin{align}\label{eq:15}
    \sup_{\bx \in \eR^{|\cH|}}&\Big|\eP\Big(\sup_{t \in \cT} \frac{\sqrt{nb_{i,l}}|\tilde \rho_{i,l}(t) - \rho_{i,l}(t)|}{\tilde \Gamma_{i,l}(t)}\leq x_{i,l},(i,l)\in\cH\Big) - \eP\Big(\Big|\frac{1}{\sqrt{n b}} \sum_{j=1}^{2\nb} \mf Z_{j}\Big|_{\infty,p}\leq \bx\Big)\Big|\notag \\
    = & O\big\{\theta^{(1)}_n + \theta^{(2)}_n\big\} = o(1). 
\end{align}
\end{theorem}

    The Gaussian approximation in \cite{bai2025time} is of the maximum-type, which is a special case of \eqref{eq:15}  with $\bx = (a,\dots,a),a\in\eR$, and it suffices for FWER control. However, FDR control requires a Gaussian approximation on hyperrectangles (derived from \cite{chernozhukov2017central}, \cite{chang2024central} and \cite{wu2024asynchronous}, with necessary modifications to adapt to our setting) to ensure the simultaneous validity of the time-varying $P$-values for each $(i,l)\in\cH$.

\begin{theorem}[Bootstrap Consistency]\label{thm2}
    Let $\bZ_\bt = (Z_{\bt,2,1},\dots,Z_{\bt,p,p-1})^\T$, where $Z_{\bt,i,l}$ is $Z_{\bt,i,l}^{(k)}$ generated in Step 6 of Algorithm \ref{alg1} in one bootstrap iteration. Under the conditions of Theorem \ref{thm1} 
    we have \begin{align}
    &\sup_{\bx \in \eR^{|\cH|}} \Big| \eP(\bZ_{\bt} \leq \bx| \cF_n) - \eP\Big(\Big| \frac{1}{\sqrt{n b}}\sum_{j=1}^{2 \nb} \mf Z_j \Big|_{\infty,p} \leq \bx\Big)\Big| = O_\eP\big\{\theta^{(2)}_n\big\} =  o_\eP(1),\label{pfthm2eq1}
    \end{align}
    where $(\mf Z_j)_{j=1}^{2\nb} \in \eR^{(n - 2\nb+1)|\cH|}$ is the Gaussian vector sequence defined in Theorem \ref{thm1}.
\end{theorem}

Theorems \ref{thm1} and \ref{thm2} show that the bootstrap procedure in
Steps 4--7 of Algorithm \ref{alg1} consistently approximates the joint
distribution of the maximum deviations of
$\tilde\rho_{i,l}(t)$ over $(i,l)\in\cH$.

We next establish the validity and power of the proposed time-varying
$P$-values. Recall that $P_{i,l}(t)$ is defined in \eqref{eq:defpval}, and $\cT = [b,1-b]$. Let $\cI_{i,l} = \{t \in\cT:(i,l)\in\cH_0(t)\}$ denote the set of time points at which the null hypothesis
$H_{i,l,t}^0$ is true. Define its interior region, after removing a boundary
band of width $b = \max_{(i,l)\in\cH} b_{i,l}$, by $\cI_{i,l}' = \{t: [t-b,t+b] \subset \cI_{i,l}\}$.

\begin{theorem}\label{thm:pvalp}
    Assume that all conditions in Theorem \ref{thm2} hold. The $P$-values $P_{i,l}(t)$ defined in \eqref{eq:defpval} satisfy: (\romannumeral 1) For any $x\in(0,1)$, \begin{equation}
        \lim_{n,B\rightarrow\infty} \sup_{(i,l)\in\cH} \eP\sth{\inf_{t\in\cI_{i,l}'} P_{i,l}(t) \le x}\le x. \notag
    \end{equation} 
    (\romannumeral 2) For any $(i,l)\in \cH, t\in\cT\backslash \cI_{i,l}$, $P_{i,l}(t)\rightarrow_\eP 0$ as $n,B\rightarrow \infty$.
\end{theorem}

Finally, we show that the time-varying BY procedure achieves the desired AuFDR control. Recall the definition of $\mathrm{AuFDR}_{r_n}$ in \eqref{eq:defFDR}, and $\ell(|\cH|) = \sum_{k=1}^{|\cH|} k^{-1} \asymp \log(|\cH|)$. Define the global true-null proportion \begin{align*}
    \pi_0^\star = \frac{|\bigcup_{t\in[0,1]}\cH_0(t)|}{|\cH|} = \frac{2|\bigcup_{t\in[0,1]}\cH_0(t)|}{p(p-1)}.
\end{align*} Let $\cC_0=\bigcup_{(i,l)\in\cH}\partial \cI_{i,l}$ denote the set of change points of $\cH_0(t)$, where $\partial \cI_{i,l}$ is the boundary of $\cI_{i,l}$.

\begin{theorem}[AuFDR Control]\label{thm3}
    Assume that the conditions of Theorem \ref{thm:pvalp} hold. Additionally assume that $p^2\{\theta_n^{(1)} + \theta_n^{(2)}\} \to 0$,  $Bp^{-2} \to \infty$, $|\cC_0|b = O(n^{-\zeta}) \to 0$ for some constant $\zeta > 0$, $r_n\rightarrow\infty$ and $r_n = o(\log n)$. Then we have 
    \begin{equation}
        \limsup_{n \to \infty} \pth{\mathrm{AuFDR}_{r_n} - \alpha \pi_0^\star }\le 0.\notag
    \end{equation}
\end{theorem}

The global true-null proportion $\pi_0^\star$ is the proportion of hypotheses that are true nulls at least once over the observation period. It appears because the AuFDR approximately corresponds to the expectation of worst pointwise FDP contribution over time, so the relevant null proportion needs to be defined over the entire observation period rather than at a fixed time point. If $\mathcal H_0(t)$ is fixed over time, $\pi_0^\star$ reduces to the usual proportion of true null hypotheses.

\begin{remark}
    In Theorem \ref{thm3}, we allow $r=r_n$ to diverge with $n$, so that $\mathrm{AuFDR}_{r_n}$ asymptotically approximates $\mathbb{E}\{\sup_t \mathrm{FDP}(t)\}$, while remaining
robust to changes in the set of true null hypotheses over time.
The convergence rate is not influenced by the choice of $r_n$, since the main argument is established on a bound of the supremum of $\mathrm{FDP}(t)$ for $t$ outside the neighborhoods of change point set $\cC_0$, whose contribution to the AuFDR can be negligible as long as $(|\cC_0|b)^{1/r_n} \rightarrow 0$.
\end{remark}


\section{Simulation Study}\label{Sec:Sim}

In this section, we evaluate the finite sample performance of Algorithm \ref{alg1} via simulations.

\subsection{Simulation Settings}

Recall the model $\bY_j=\bmu(t_j)+\bG(t_j,\cF_j), j\in[n]$. In the following, we present details of our simulation setting. Let $\sth{\eta_{j,i},j\in[n],i\in[p]}$ be a collection of i.i.d.\ random variables and define $\boldsymbol{\eta}_j = (\eta_{j,1}, \dots, \eta_{j,p})^\top$, where each $\eta_{j,i}$ follows either a standard Gaussian or a standard Laplace distribution. Let $\bI_p$ denote the $p$-dimensional identity matrix, $\bJ_p$ denote $p$-dimensional matrix with each entry being $1$.

\textbf{Case 1} ($p = 6$):  Let $\bA(t) = f(t)\bI_6$, with $f(t) = 0.4 - 0.1 (t - 1/2)^2$. For the error process, we set $\mathbf{G}(t, \mathcal{F}_j) = \mathbf{A}(t) \mathbf{G}(t, \mathcal{F}_{j-1}) + (4\bI_6/5 + \bI_2\otimes\bJ_3/5) \boldsymbol{\eta}_j$. The mean functions are defined as
\begin{align*}
    &\mu_i(t) =(0.3+0.4t)\mathbf{1}(t\le a_{i,1})+(0.7-0.4t)\mathbf{1}(a_{i,1}< t\le a_{i,2})+(0.2+0.4t)\mathbf{1}(t > a_{i,2}),
\end{align*}
where for $i = 1,4$, $a_{i,1} = 0.35$ and $a_{i,2} = 0.65$; for $i = 2,5$, $a_{i,1} = 0.5$ and $a_{i,2} = 0.8$; and for $i = 3,6$, $a_{i,1} = 0.65$ and $a_{i,2} = 0.95$. The global true-null proportion is $\pi_0^\star = 0.6$.

\textbf{Case 2} ($p = 9$): Similar to Case 1, the error process is  $\mathbf{G}(t, \mathcal{F}_j) = \mathbf{A}(t) \mathbf{G}(t, \mathcal{F}_{j-1}) + (4\bI_9/5 + \bI_3\otimes\bJ_3/5) \boldsymbol{\eta}_j$, where $\bA(t) = f(t)\bI_9$. For the mean function, let $\mu_i(t)$, $i \in [6]$, be the same as in Case~1, and set $\mu_7(t) = \mu_1(t)$, $\mu_8(t) = \mu_2(t)$, and $\mu_9(t) = \mu_3(t)$. The global true-null proportion is $\pi_0^\star = 0.75$.

\textbf{Case 3} ($p = 6$): This case is designed to evaluate the method when the set of true null hypotheses changes over time. Let $\boldsymbol{\eta}_{j}^{(1)}=(\eta_{j,1},\eta_{j,2},\eta_{j,3})^\top$ and $\boldsymbol{\eta}_{j}^{(2)}=(\eta_{j,4},\eta_{j,5},\eta_{j,6})^\top$. For $\boldsymbol{\epsilon}_j^{(1)} = (\epsilon_{j,1},\epsilon_{j,2},\epsilon_{j,3})^\top = \bG^{(1)}(j/n,\cF_j)$, it follows
\begin{align*}
    \mathbf{G}^{(1)}(t,\mathcal{F}_j)
    = f(t)\mathbf{G}^{(1)}(t,\mathcal{F}_{j-1})
    + \mathbf{B}_0(t)\boldsymbol{\eta}_{j}^{(1)},
    \qquad
    \mathbf{B}_0(t)=\frac{1}{5}
    \begin{pmatrix}
        5 & g(t) & g(t)\\
        g(t) & 5 & g(t)\\
        g(t) & g(t) & 5
    \end{pmatrix},
\end{align*}
where $f(t)=0.4-0.1(t-1/2)^2$ and
\begin{align*}
    g(t)=
    \begin{cases}
        1, & t<0.45,\\
        1-(t-0.45)/0.25, & 0.45\le t<0.7,\\
        0, & t\ge 0.7.
    \end{cases}
\end{align*}
For $\boldsymbol{\epsilon}_j^{(2)} = (\epsilon_{j,4},\epsilon_{j,5},\epsilon_{j,6})^\top = \bG^{(2)}(j/n,\cF_j)$, it follows $\bG^{(2)}(t,\cF_j) = \bA(t)\bG^{(2)}(t,\cF_{j-1}) + (4\bI_3/5 + \bJ_3) \bfeta_j^{(2)}$, where $\bA(t) = f(t)\bI_3$ with $f(t)$ defined in Case~1. The mean functions are the same as in Case~1. Thus, the correlations among variables $1,2,3$ gradually vanish and become null after $t=0.7$, while the correlations among variables $4,5,6$ remain nonzero. The global true-null proportion is $\pi_0^\star = 0.8$.

\textbf{Case 4} ($p = 9$): For $\boldsymbol{\epsilon}_j^{(1)} = (\epsilon_{j,1},\epsilon_{j,2},\epsilon_{j,3})^\top$ and $\boldsymbol{\epsilon}_j^{(2)} = (\epsilon_{j,4},\epsilon_{j,5},\epsilon_{j,6})^\top$, they follow the same distribution as in Case~3. For $\boldsymbol{\epsilon}_j^{(3)} = (\epsilon_{j,7},\epsilon_{j,8},\epsilon_{j,9})^\top = \bG^{(3)}(j/n,\cF_j)$, it follows $\bG^{(3)}(t,\cF_j) = \bA(t)\bG^{(3)}(t,\cF_{j-1}) + (4\bI_3/5 + \bJ_3) \bfeta_j^{(3)}$, where $\bfeta^{(3)}_j = (\eta_{j,7},\eta_{j,8},\eta_{j,9})^\top$,  $\bA(t) = f(t)\bI_3$ with $f(t)$ defined in Case~1. The mean functions are the same as in Case~2. The global true-null proportion is $\pi_0^\star = 5/6$.

\subsection{Simulation Results}

For Cases~1--2, we set sample sizes $n \in \{600, 900, 1500\}$. For Cases~3--4, which correspond to the time-varying $\cH_0(t)$ setting, we use $n \in \{900, 1500, 2400\}$. The lag parameter is set as $h = \lceil \log n \rceil$. For all cases, the BY procedure is applied at target levels $\alpha \in\{0.05,0.1\}$. All results are reported as averages over 100 Monte Carlo repetitions.

To mitigate the boundary effects, we evaluate the asymptotically uniform FDP (AuFDP) over the truncated interval $[0.15, 0.85]$, defined as \begin{align*}
    \mathrm{AuFDP}_{r_n} = \left[ \int_{0.15}^{0.85} 0.7^{-1} \left\{ \fdp(t) \right\}^{r_n} \mathrm{d}t \right]^{1/r_n}.
\end{align*}
In the report of simulation results, we take $r_n = 2(\log n)^{1/2}$. The false negative proportion (FNP) at time $t$ is denoted as $\mathrm{FNP}(t)$, and we report the average FNP across $t\in[0.15,0.85]$.

Tables~\ref{tab:summ} and \ref{tab:summtv} summarize the simulation results for Cases~1--2 and Cases~3--4 respectively, and we have the following observations.

\begin{itemize}
    \item In all simulation scenarios, increasing $n$ substantially reduces average FNPs. For sufficiently large sample sizes, the values of $\mathrm{AuFDP}_{r_n}$ are generally close to or below the theoretical upper bound in Theorem~\ref{thm3}, supporting our theoretical analysis. Moreover, Table~\ref{tab:summtv} suggests that our procedure is robust to temporal changes in $\mathcal H_0(t)$, highlighting the advantage of AuFDR over the supremum FDR criterion.
    \item The Laplacian innovations are more challenging than the Gaussian innovations and typically lead to larger average FNPs, but the false-negative rates still decrease with $n$ and the $\mathrm{AuFDP}_{r_n}$ remains close to or below the theoretical upper bound for sufficient sample sizes $n$.
\end{itemize}

\begin{table}[t]
    \centering
    \caption{$\mathrm{AuFDP}_{r_n}$'s and average FNPs for different levels $\alpha$, with $r_n = 2(\log n)^{1/2}$.}
    \label{tab:summ}
    \small 
    \begin{tabular}{cccc cc cc}
        \toprule
        \multirow{2}{*}{Tail} & {\multirow{2}{*}{Case}} & \multirow{2}{*}{$p$} & \multirow{2}{*}{$n$} & \multicolumn{2}{c}{$\mathrm{AuFDP}_{r_n}$} & \multicolumn{2}{c}{Average FNP} \\
        \cmidrule(lr){5-6} \cmidrule(lr){7-8}
        & & & & $\alpha=0.05$ & $\alpha=0.10$ & $\alpha=0.05$ & $\alpha=0.10$ \\
        \midrule
        \multirow{6}{*}{Gaussian} & {\multirow{3}{*}{1}} & \multirow{3}{*}{6} 
              & 600  & 0.037 & 0.057 & 0.087 & 0.066 \\
            & & & 900  & 0.024 & 0.043 & 0.032 & 0.024 \\
            & & & 1500 & 0.031 & 0.042 & 0.005 & 0.003 \\
        \cmidrule(lr){2-8}
            & {\multirow{3}{*}{2}} & \multirow{3}{*}{9} 
              & 600  & 0.045 & 0.057 & 0.110 & 0.085 \\
            & & & 900  & 0.041 & 0.056 & 0.055 & 0.037 \\
            & & & 1500 & 0.043 & 0.060 & 0.008 & 0.004 \\
        \midrule
        \multirow{6}{*}{Laplacian} & {\multirow{3}{*}{1}} & \multirow{3}{*}{6} 
              & 600  & 0.038 & 0.047 & 0.112 & 0.085 \\
            & & & 900  & 0.017 & 0.023 & 0.068 & 0.048 \\
            & & & 1500 & 0.020 & 0.025 & 0.029 & 0.018 \\
        \cmidrule(lr){2-8}
            & {\multirow{3}{*}{2}} & \multirow{3}{*}{9} 
              & 600  & 0.042 & 0.054 & 0.150 & 0.119 \\
            & & & 900  & 0.024 & 0.035 & 0.100 & 0.075 \\
            & & & 1500 & 0.019 & 0.025 & 0.050 & 0.033 \\
        \bottomrule
    \end{tabular}
\end{table}

\begin{table}[t]
    \centering
    \caption{$\mathrm{AuFDP}_{r_n}$'s and average FNPs for different levels $\alpha$, with $r_n = 2(\log n)^{1/2}$, under the time-varying $\cH_0(t)$ settings.}
    \label{tab:summtv}
    \small
    \begin{tabular}{cccc cc cc}
        \toprule
        \multirow{2}{*}{Tail} & \multirow{2}{*}{Case} & \multirow{2}{*}{$p$} & \multirow{2}{*}{$n$} & \multicolumn{2}{c}{$\mathrm{AuFDP}_{r_n}$} & \multicolumn{2}{c}{Average FNP} \\
        \cmidrule(lr){5-6} \cmidrule(lr){7-8}
        & & & & $\alpha=0.05$ & $\alpha=0.10$ & $\alpha=0.05$ & $\alpha=0.10$ \\
        \midrule
        \multirow{6}{*}{Gaussian} & \multirow{3}{*}{3} & \multirow{3}{*}{6}
            & 900  & 0.057 & 0.078 & 0.118 & 0.102 \\
            & & & 1500 & 0.056 & 0.067 & 0.088 & 0.079 \\
            & & & 2400 & 0.044 & 0.061 & 0.067 & 0.061 \\
        \cmidrule(lr){2-8}
            & \multirow{3}{*}{4} & \multirow{3}{*}{9}
            & 900  & 0.055 & 0.073 & 0.116 & 0.094 \\
            & & & 1500 & 0.053 & 0.073 & 0.069 & 0.060 \\
            & & & 2400 & 0.041 & 0.052 & 0.051 & 0.046 \\
        \midrule
        \multirow{6}{*}{Laplacian} & \multirow{3}{*}{3} & \multirow{3}{*}{6}
            & 900  & 0.037 & 0.052 & 0.166 & 0.142 \\
            & & & 1500 & 0.034 & 0.042 & 0.124 & 0.106 \\
            & & & 2400 & 0.027 & 0.031 & 0.104 & 0.092 \\
        \cmidrule(lr){2-8}
            & \multirow{3}{*}{4} & \multirow{3}{*}{9}
            & 900  & 0.031 & 0.047 & 0.168 & 0.142 \\
            & & & 1500 & 0.027 & 0.035 & 0.117 & 0.097 \\
            & & & 2400 & 0.015 & 0.022 & 0.093 & 0.075 \\
        \bottomrule
    \end{tabular}
\end{table}

\subsection{Additional Simulations}\label{Sec:Base}\label{sec:Sensi}
For brevity, we present some additional simulations in Appendix~\ref{app:SecC}. Appendix~\ref{app:SecC1} reports baseline comparisons of the proposed method with two competing approaches: the moving-window method in \cite{masuda2025introduction} and the time-varying adaptation of multiple-testing procedure of \cite{cai2016large}. Appendix~\ref{app:SecC2} reports the sensitivity analysis for the bandwidths selected by the data-driven procedure in Section \ref{sec.3.4} and for the lag parameter $h$.

\section{Real Data Analysis}\label{sec:RDA}

\subsection{EEG Data}

    Electroencephalography (EEG) measures the electrical activity of the brain using sensors placed on the scalp and facilitates many neuroscience studies. In this section, we analyze an EEG dataset, in order to explore the relationship between alcoholism and brain activity. Details regarding data collection procedures can be found in \cite{zhang1997electrophysiological}. While the original analysis focuses on amplitudes of event-related potentials (ERPs), we adopt an alternative perspective by inferring time-varying correlations of the ERPs.

    The data in our analysis are available at \url{https://archive.ics.uci.edu/dataset/121/eeg+database}, and we focus on the SMNI\_CMI\_TRAIN dataset. It comprises 10 subjects from the alcoholic group and 10 subjects from the control group. Each subject performed 10 image recognition tasks when a single image was shown, with each task lasting one second, resulting in 10 segments per subject. Following the methodology in \cite{zhang1997electrophysiological}, the ERPs for each subject were averaged across these segments to obtain mean ERPs. ERPs were recorded using $p = 64$ electrodes, and each ERP segment consists of $n = 256$ time points, corresponding to a sampling rate of 256 Hz over one second. For each subject, we run Algorithm~\ref{alg1} to obtain the time-varying rejection set $\cR(t)$. The target BY level is set as $\alpha = 0.2$, equivalently using the adjusted nominal level $\alpha/\ell(|\cH|)=0.0244$. We further visualize the result via constructing the time-varying correlation network $\mathcal{G}(t) = \{\mathcal{V}, \mathcal{E}(t)\}$, with node set $\cV = [p]$ and edge set $\cE(t) = \sth{(i,l): i,l\in[p], (i,l)~\mbox{or}~(l,i) \in \cR(t)}$.

\begin{figure}[htbp]
    \centering
    \includegraphics[width=1\linewidth]{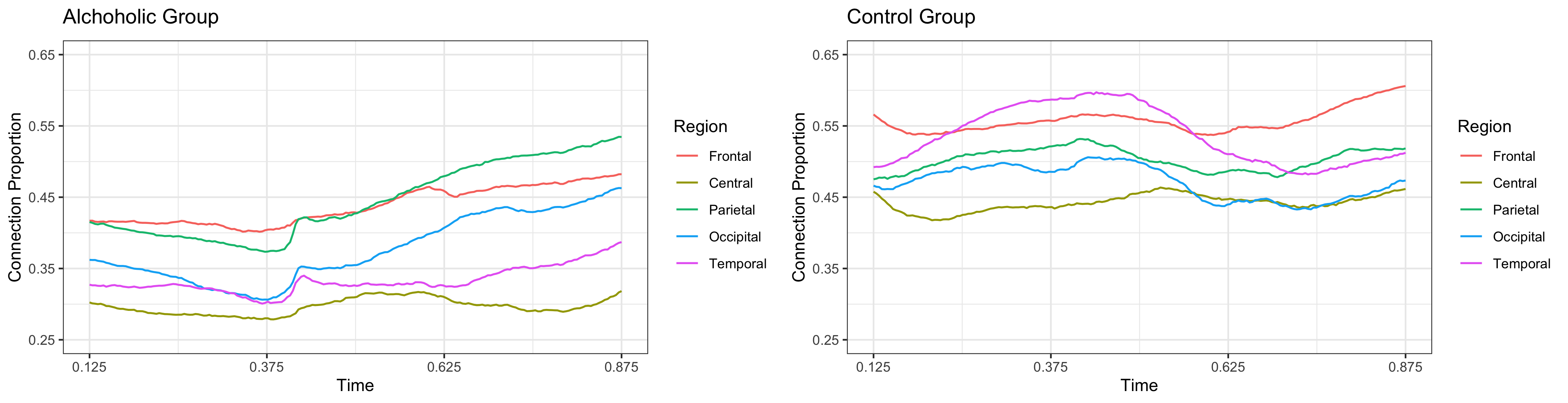}
    \caption{Time-varying connection proportions (averaged across subjects in each group).}
    \label{fig:EEG}
\end{figure}

    The 64 scalp electrodes can be grouped into five anatomical regions: frontal, central, parietal, occipital, and temporal. Let $\mathcal{V}_1, \dots, \mathcal{V}_5$ denote the sets of nodes corresponding to these five anatomical regions, and $\cV$ denote the set of all nodes. We define the connection proportion for region $k\in[5]$ at time $t$ as
    \begin{align*}
    \zeta_k(t) = |\mathcal{V}_k|^{-1} |\mathcal{V}|^{-1} \sum_{v_1 \in \mathcal{V}_k} \sum_{v_2 \in \mathcal{V}} \mathbf{1}\{(v_1,v_2)\in\cE(t)\}.
    \end{align*}
    Figure~\ref{fig:EEG} displays the time-varying connection proportions for each region (averaged across 10 subjects in each group). We have the following observations:

    (\romannumeral 1) The connection proportions for each region exhibit temporal variations with upward trends, reflecting changes in brain activity in response to the image stimuli presented to the subjects.

    (\romannumeral 2) Control subjects demonstrate visibly higher average connection proportions in all brain regions. Similarly, \cite{zhang1997electrophysiological} reported stronger brain activity in the temporal, occipital, and frontal regions of control subjects, suggesting that long-term alcohol abuse may impair functioning in these regions. While their analysis focused on amplitudes, our results additionally highlight group differences in time-varying correlations.

\subsection{Economic Data}

    The interconnectedness among economic data plays a crucial role in quantifying risks within financial portfolios, as well as systemic and macroeconomic risks. For an introduction, see \cite{acemoglu2012network}, \cite{puliga2014credit}, and \cite{guo2018development}, among others. In this section, we infer the time-varying correlation of Daily WRDS World Indices, which are market-capitalization-weighted indices with dividends at daily frequency (for details, see \url{https://wrds-www.wharton.upenn.edu/pages/get-data/world-indices-wrds/}).

    Specifically, we analyze the weekly averages of the absolute values of the Daily Country Return with Dividends (PORTRET) from June 1, 2006, to June 1, 2022, which can serve as measures of volatility and risk. We denote the weekly average of the absolute value of PORTRET at time \( j \) for region \( i \) by \( Y_{j,i} \). There are $n=819$ time points for $p = 6$ regions encompassing France, Germany, the UK, China, Japan, and Hong Kong. Based on these observations, we apply Algorithm \ref{alg1} and set the target BY level as $\alpha = 0.2$, equivalently using the adjusted nominal level $\alpha/\ell(|\cH|)=0.0603$. For visualization, the time-varying networks are presented in Figure \ref{fig:RDA}. For brevity, the figure displays only the time-varying networks at the first week of June in each year. We restrict the visualization to time points lying away from the boundary and only report the networks for 2011--2018. From the figure, we have the following observations:

\begin{figure}[h]
    \centering
    \includegraphics[width=0.8\linewidth]{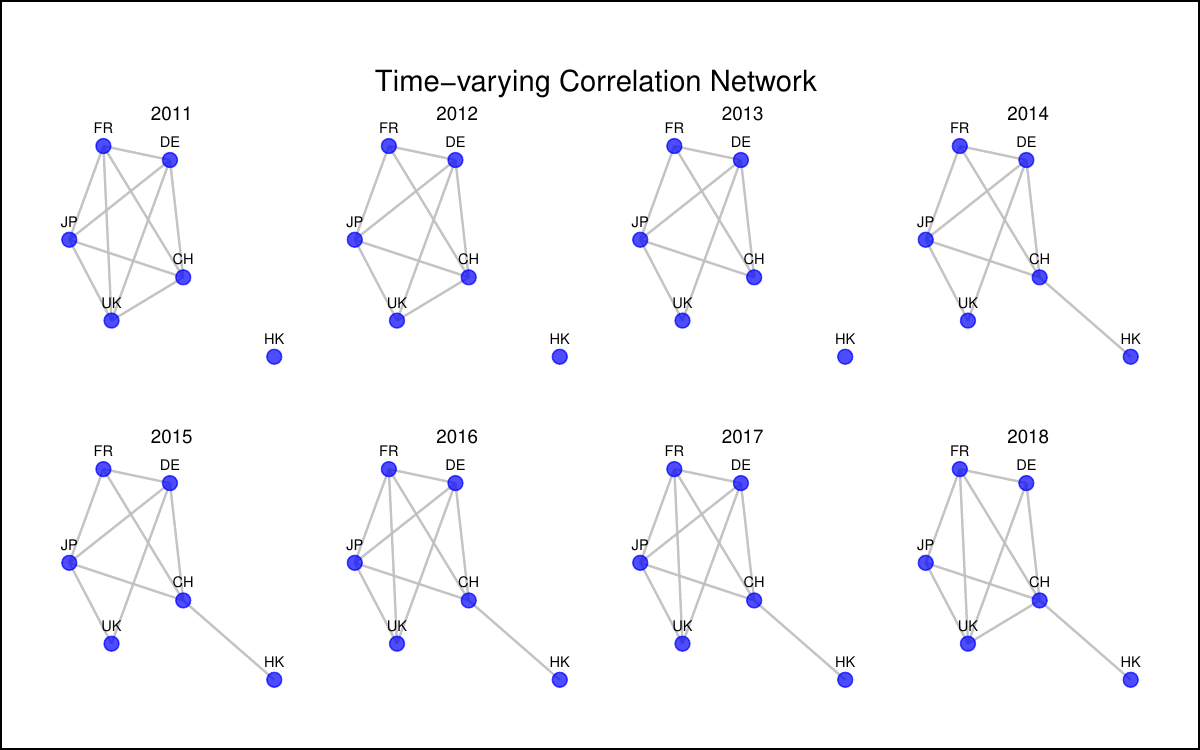}
    \caption{Correlation network of WRDS data for selected years 2011--2018.}
    \label{fig:RDA}
\end{figure}

    (\romannumeral 1) The inference results are broadly consistent with known regional/economic linkages. For example,  we identify persistent correlations over time between France and Germany, China and Japan, and the UK and Germany, consistent with their strong economic relationships.
    
    (\romannumeral 2) The results also capture substantial temporal variation in correlation. For example, correlations of volatility between Hong Kong and other regions are not statistically significant after the BY adjustment during 2011--2013. This period corresponds to a sharp increase of volatility of the stock market in Hong Kong \citep{hkma2012economic}. Additionally, in 2012, the interconnectedness among European countries experienced an obvious shift, driven by the European sovereign debt crisis.

    To demonstrate the effectiveness of our method in capturing correlations, we additionally construct a semi-synthetic dataset. We introduce a synthesized country whose weekly average of the absolute value of PORTRET at time $t$ is defined as follows. Let $\{\widetilde{Y}_j,j\in[819]\}$ be obtained by randomly sampling without replacement from $\sth{Y_{j,i},i\in[6],j\in[819]}$, and $\{Y_{j,1},j\in[819]\}$ represents the data of China. We construct the data of the synthesized country as $\sth{Y_{j,7},j\in[819]}$, where \begin{align*}
    Y_{j,7} = \begin{cases}
        \widetilde{Y}_j ~~\mathrm{if}~~ j\le n/3,\\
        (7/3-4j/n)\widetilde{Y}_j + (4j/n-4/3) Y_{j,1} ~~\mathrm{if}~~ n/3 < j\le n/2,\\
        \widetilde{Y}_j/3 + 2Y_{j,1}/3~~\mathrm{if}~~ j > n/2,\\
    \end{cases}
\end{align*}

The resulting semi-synthetic data are denoted by 
$\{Y_{j,i}:j\in[819],i\in[7]\}$. 
We then apply Algorithm~\ref{alg1} to estimate the time-varying correlation network. 
The correlation signal associated with the synthetic country is successfully detected, with an absolute discrepancy of only $0.07$ between the detected time and the true signal time. This implies that our method is capable of detecting time-varying signals accurately and efficiently.

\section{Conclusion}\label{sec:conclusion}

This paper studies simultaneous inference for high-dimensional locally stationary time series, with a focus on time-varying correlations. We propose the notion of AuFDR, which is defined as the expectation of $L_{r_n}$-norms of the time-varying FDP, with $r_n\rightarrow\infty$ as sample size $n\rightarrow\infty$. This criterion provides a principled compromise between pointwise FDR control, which fails to ensure reliability over an entire time interval, and the supremum-type criterion, which can be overly sensitive to abrupt local changes in the signal.

Building on this criterion, we develop a general inferential framework for detecting time-varying correlations. The procedure combines robust difference-based estimation with a multiplier bootstrap to generate simultaneously valid $P$-values,  followed by  a time-varying BY step for AuFDR control under arbitrary dependence. The method is robust to non-Gaussianity, nonlinear dependence, and jumps in means, features common in modern time series data.

Several promising directions remain for future research. First, while this paper focuses on marginal correlations, many applications require inference on conditional dependence structures, such as time-varying partial correlations or graphical models. Extending the AuFDR framework to these settings is an important direction. Second, although the time-varying BY procedure achieves uniform control in our framework, more refined FDR procedures, including adaptive methods, may further improve power while preserving AuFDR guarantees. Third, the AuFDR framework, as a general error-control criterion, may also be useful in other inference problems involving time-varying structures of locally stationary time series.

\section{Disclosure Statement}\label{disclosure-statement}

The authors have no conflicts of interest to declare.

\section{Data Availability Statement}\label{data-availability-statement}

The brain EEG data is openly available in EEG Database at \url{http://doi.org/10.24432/C5TS3D}. The financial data were obtained from the Daily WRDS World Indices database through Wharton Research Data Services (WRDS), University of Pennsylvania (\url{https://wrds-www.wharton.upenn.edu/pages/get-data/world-indices-wrds/}). The raw WRDS data are only available from WRDS to users with appropriate institutional subscription and permission.

\bibliography{bibliography}

\clearpage

\appendix

\begin{center}

{\large\bf SUPPLEMENTARY MATERIAL}

\end{center}

\section{Technical Appendix}\label{app:SecD}

We first provide some additional notations used throughout the Appendix. Let $\rightarrow_\eP$ denote convergence in probability. We use $a \vee b, a\wedge b$ to denote $\max\{a, b\}$ and $\min\sth{a,b}$. For an event $A$, let $\bar{A}$ denote its complement. Throughout the Appendix, $|\cH|=p(p-1)/2$ is the number of hypotheses. Let $\mf 1(\cdot)$ denote the indicator function. For two positive sequences $a_n,b_n$, write $a_n\lesssim b_n$ if there exists a constant $C$ and $N\in\eN$ such that $a_n\le C b_n$ for all $n\ge N$. Write $a_n\asymp b_n$ if $a_n\lesssim b_n$ and $b_n\lesssim a_n$. Define $\epsilon^{(h)}_{j,i} = \epsilon_{j,i} - \epsilon_{j-h,i},~j\in[n],~i\in[p]$ as $h$-th order differenced error. We use $C, C'$ to denote absolute constants whose values may change from line to line. Constants with a symbolic subscript, such as $C_{\star},C'_{\star},$ are used to indicate that their values depend only on the subscript, and may also vary from line to line.

\subsection{Proof of Theorem \ref{thm1}}

Recall that $b = \max_{i,l\in[p]} b_{i,l},\cT = [b,1-b]$ and further define $\cT_n = \{\nb, \nb+1,\dots, n-\nb\}$. Recall the definition of $\vartheta_{i,l}(t),\Xi_{j,i,l}$ as in \eqref{eq8} and \eqref{eq:Xidiscrete}. Define the $\cF_n$-measurable event \begin{align*}
    B_n' = \sth{\max_{i,l\in[p]}\sup_{t\in\cT}\ath{\tilde{\gamma}_{i,l}(t)-\gamma_{i,l}(t)} > f_nq_n'},
\end{align*}
where $f_n=(nb)^{-1/2}(p^2/b)^{1/q}$ and $q_n'$ is a positive sequence such that $q_n'\rightarrow\infty$ and $f_nq_n'\rightarrow 0$. Recall that $n^\phi$ is the upper bound of number of abrupt jumps as in Assumption \ref{Ass:error}(\romannumeral 6). By Lemma \ref{lem:lowodker}, we have \begin{align}
    \left\| \max_{ i,l\in[p]}\max_{ j\in \cT_n}\left|\tilde \rho_{i,l}(t_j) - \rho_{i,l}(t_j) - \vartheta_{i,l}(t_j)\right| \mathbf 1(\bar B^{\prime}_n)\right\|_q & = O\sth{(p^2/b)^{1/q} (hn^{\phi - 1}b^{-1}  + b^{2}+n^{-1/2}h)}\notag \\
    &= O\sth{(p^2/b)^{1/q} (n^{\phi - 1}b^{-1}h  + b^{2})}:= O(c_n),\notag\\
    \eP(B^{\prime}_n) & =O\sth{\pth{q_n'}^{-q}}.\label{pfthm1eq2}
\end{align}
Recall the definition of $\bar{\mf \Xi}^{\cG}_i$ and the definition of norm $|\cdot|_{\infty,p}$ in \eqref{eq6}. Lemma \ref{lm2} implies \begin{align*}
    \sup_{\bx\in\eR^{{|\cH|}}}\ath{\eP\pth{\ath{\frac{1}{\sqrt{nb}}\sum_{i=1}^{2\lceil nb \rceil}\bar{\mf \Xi}^{\cG}_i}_{\infty,p} \le \bx} - \eP\pth{\ath{\frac{1}{\sqrt{nb}}\sum_{i=1}^{2\lceil nb \rceil}\mf Z_i}_{\infty,p} \le \bx}} = O\sth{(nb)^{-1/8}\pth{\log n}^{4}}.
\end{align*}
By Lemma \ref{lm1} and \eqref{pfthm1eq2}, for any $\delta > 0$, \begin{align}
    &\sup_{\bx\in\eR^{{|\cH|}}}\ath{\eP\pth{ \max_{j\in\cT_n}\sqrt{nb_{i,l}}|\tilde{\rho}_{i,l}(t_j)-\rho_{i,l}(t_j)|/\Gamma_{i,l}(t_j) \le x_{i,l},i,l\in[p]} - \eP\pth{\ath{\frac{1}{\sqrt{nb}}\sum_{i=1}^{2\lceil nb \rceil}\mf Z_i}_{\infty,p} \le \bx}} \notag\\
    &= O\sth{(nb)^{-1/8}\pth{\log n}^{4}+\delta\sqrt{\log (n p)}+(q_n')^{-q} + (\sqrt{nb}c_n/\delta)^q}.\label{pfthm1eq1}
\end{align}
Solving $(\sqrt{nb}c_n/\delta)^q = \delta = (q_n^{\prime})^{-q}$, we have $\delta = (\sqrt{nb}c_n)^{q/(q+1)}, q_n^{\prime} = (\sqrt{nb}c_n)^{-1/(q+1)}$. From Assumption \ref{Ass:bdrate}, $c_n = o((nb)^{-1/2})$ and $q_n^{\prime} \to \infty$. Under the bandwidth condition $p^2/(n^{q/2+\phi}b^{q/2+1}h+b^{q/2+4}n^{1+q/2})\rightarrow 0$, we have $f_nq_n'\to 0$.
Note that $c_n  = o\sth{(nb)^{-1/2}} = o(b^{-1})$. From Equation (41) of \cite{bai2025time}, we have \begin{align}\label{pfthm1eq3}
    &\left\| \max_{i,l\in[p]} \sqrt{nb_{i,l}} \sup_{|t_j- t| \leq n^{-1}, j \in \cT_n, t\in[0,1]}\left |\frac{\tilde \rho_{i,l}(t) - \rho_{i,l}(t)}{\Gamma_{i,l}(t)} - \frac{ \tilde \rho_{i,l}(t_j) - \rho_{i,l}(t_j)}{ \Gamma_{i,l}(t_j)}\right|\mf 1(\bar B^{\prime}_n)\right\|_q\notag \\
    &= O\sth{p^{2/q} (nb)^{-1}}.
\end{align}
By Lemma \ref{lm1}, the above inequality and \eqref{pfthm1eq1}, we have for any $\delta' > 0$, \begin{align*}
    &\sup_{\bx\in\eR^{{|\cH|}}}\ath{\eP\pth{ \max_{t\in\cT}\sqrt{nb_{i,l}}|\tilde{\rho}_{i,l}(t)-\rho_{i,l}(t)|/\Gamma_{i,l}(t) \le x_{i,l},i,l\in[p]} - \eP\pth{\ath{\frac{1}{\sqrt{nb}}\sum_{i=1}^{2\lceil nb \rceil}\mf Z_i}_{\infty,p} \le x}} \notag\\
    &= O\sth{(nb)^{-1/8}\pth{\log n}^{4}+ (\sqrt{nb}c_n)^{q/(q+1)}\sqrt{\log (np)} + \delta'\sqrt{\log (np)}+(q_n')^{-q} + \pth{\frac{p^{2/q}}{nb\delta'}}^q}.
\end{align*}
In the previous argument we have already set $q_n^{\prime} = (\sqrt{nb}c_n)^{-1/(q+1)}$. In this step, set $\delta' = \sth{p^{2/q}/(nb\delta')}^q$, which gives $\delta' = p^{2/(q+1)}(nb)^{-q/(q+1)}$. We have \begin{align*}
    &\sup_{\bx\in\eR^{|\cH|}}\ath{\eP\pth{ \max_{t\in\cT}\sqrt{nb_{i,l}}|\tilde{\rho}_{i,l}(t)-\rho_{i,l}(t)|/\Gamma_{i,l}(t) \le x_{i,l},i,l\in[p]} - \eP\pth{\ath{\frac{1}{\sqrt{nb}}\sum_{i=1}^{2\lceil nb \rceil}\mf Z_i}_{\infty,p} \le x}}\\
    &= O\sth{(nb)^{-1/8}\pth{\log n}^{4}+ \sth{(\sqrt{nb}c_n)^{q/(q+1)}+p^{2/(q+1)}(nb)^{-q/(q+1)}}\sqrt{\log (np)} }.
\end{align*} 
Let $g^{\prime}_n =p^{2/q} (\sqrt{m/(n \eta^{2})}+1/m+\eta  +\sqrt{m/(nb)}(mb/n)^{-1/2q})$, $g_n = p^{2/q}\big(\sqrt{\frac{m}{n \eta^{2}}}+m^{-1}+\eta  +(\frac{m}{nb})^{1/2}(\frac{mb}{n})^{-1/(2q)}+ (nb)^{-1/2}b^{-1/q}\big) + w^{3/2}/n + h n^{\phi} /\sqrt{w}$.
By Lemma \ref{lm1} and \ref{lem:plugingap}, for any $\delta^\diamond > 0$, \begin{align*}
    &\sup_{\bx\in\eR^{|\cH|}}\ath{\eP\pth{ \max_{t\in\cT}\sqrt{nb_{i,l}}|\tilde{\rho}_{i,l}(t)-\rho_{i,l}(t)|/\tilde\Gamma_{i,l}(t) \le x_{i,l},i,l\in[p]} - \eP\pth{\ath{\frac{1}{\sqrt{nb}}\sum_{i=1}^{2\lceil nb \rceil}\mf Z_i}_{\infty,p} \le x}}\\
    &= O\sth{(nb)^{-1/8}\pth{\log n}^{4}+ \sth{(\sqrt{nb}c_n)^{q/(q+1)}+p^{2/(q+1)}(nb)^{-q/(q+1)}}\sqrt{\log (np)}}\\
    & + O\sth{\pth{g_n(np^2)^{1/q}}^{q/(q+2)} + (\sqrt{nb}c_n)^{q/(q+1)} +(\delta^{\diamond})^{-q/2}\sth{p^2b^{-1/2}(g_n')^{q/2}} +\delta^\diamond\sqrt{\log (np)}}
\end{align*}
Set $(\delta^{\diamond})^{-q/2}\sth{p^2b^{-1/2}(g'_n)^{q/2}} = \delta^\diamond$, we have $\delta^\diamond = p^{4/(q+2)} b^{-1/(2+q)}(g_n')^{q/(2+q)}$. By $g_n \ge g_n'$, \begin{align*}
    &\pth{g_n(np^4)^{1/q}}^{q/(q+2)}/p^{4/(q+2)} b^{-1/(2+q)}(g_n)^{q/(2+q)} =n^{1/(q+2)}b^{1/(q+2)}p^{-2/(q+2)} \rightarrow \infty,
\end{align*}
we have $\delta^\diamond\sqrt{\log (np)} + \pth{g_n(np^2)^{1/q}}^{q/(q+2)} = O\sth{\pth{g_n(np^4)^{1/q}}^{q/(q+2)}\sqrt{\log (np)}}$. Noticing that $\log (np) = O\pth{\log n}$ since $p = O(n^{\iota})$, this gives \begin{align*}
    &\sup_{\bx\in\eR^{|\cH|}}\ath{\eP\pth{ \max_{t\in\cT}\sqrt{nb_{i,l}}|\tilde{\rho}_{i,l}(t)-\rho_{i,l}(t)|/\tilde\Gamma_{i,l}(t) \le x_{i,l},i,l\in[p]} - \eP\pth{\ath{\frac{1}{\sqrt{nb}}\sum_{i=1}^{2\lceil nb \rceil}\mf Z_i}_{\infty,p} \le x}}\\
    &= O\sth{(nb)^{-1/8}\pth{\log n}^{4}+ \sth{(\sqrt{nb}c_n)^{q/(q+1)}+p^{2/(q+1)}(nb)^{-q/(q+1)}}\sqrt{\log n}}\\
    & + O\sth{\pth{g_n(np^4)^{1/q}}^{q/(q+2)}\log n } = O(\theta_n^{(1)} + \theta_n^{(2)}) = o(1).
\end{align*} 
In the last line we use $\theta^{(1)}, \theta^{(2)} \rightarrow 0$ as in Assumption \ref{Ass:bdrate}. Therefore we finish the proof of Theorem \ref{thm1}.

\subsection{Proof of Theorem \ref{thm2}}

    Let $Z_{\bt,i,l}$ denote $Z_{\bt,i,l}^{(k)}$ in one iteration of Algorithm \ref{alg1}. 
    Let $\bZ^\diamond = (\bZ_{2,1}^\diamond,\dots,\bZ_{p,p-1}^\diamond)^\T$, with $\bZ_{i,l}^\diamond = (\bZ_{1,i,l}^\diamond,\dots,\bZ_{n-2\nb+1,i,l}^\diamond)^\T,i,l\in[p]$, where \begin{align*}
        Z^{\diamond}_{a,i,l} =  (2w\nb)^{-1/2}\sum_{j = w}^{2\nb - w} \hat{\tilde S}_{a-1, j, i, l} R_{a+j-1}, \quad a\in[n-2\nb +1], i,l\in[p].
    \end{align*}
    Here $\bZ^\diamond$ is the set of multiplier Gaussian variables across all dimension $i,l\in[p]$ and across all time $a\in[n-2\nb + 1]$, $\bZ^\diamond_{i,l}$ is the set of multiplier Gaussian variables for the dimension $i,l\in[p]$ across all time $a\in[n-2\nb + 1]$, and $Z^{\diamond}_{a,i,l}$ is the multiplier Gaussian variable for dimension $i,l\in[p]$ at time $a$. It follows that $Z_{\bt,i,l} = |\bZ_{i,l}^\diamond|_\infty$. Recall the definition of $\bar \bXi_{j,s}^\cG, \bar \Xi_{j,s,i,l}$ before Theorem \ref{thm1}. Define \begin{align}
       \tilde S_{a-1, j,i,l} = \sum_{i=j-w+1}^{j} \bar{\Xi}_{i + (a-1), \nb + (a-1), i, l} - \sum_{i=j+1}^{j+w}\bar{\Xi}_{i + (a-1), \nb + (a-1), i, l},\notag
    \end{align}
    and $\bZ^{\dagger}$ by substituting $\hat{\tilde {S}}_{(a-1), j,i,l}$ in $\bZ^{\diamond}$ by $\tilde{S}_{(a-1), j,i,l}$. Recall the definition of $\bZ_j$ in Theorem \ref{thm1}.  Define $\sigma^{\bZ^{\diamond}}_{k_1,k_2,i_1,l_1,i_2,l_2} = \eE\pth{Z^\diamond_{k_1,i_1,l_1}Z^\diamond_{k_2,i_2,l_2}\mid\cF_n}, \sigma^{\bZ^{\dagger}}_{k_1,k_2,i_1,l_1,i_2,l_2} = \eE\pth{Z^\dagger_{k_1,i_1,l_1}Z^\dagger_{k_2,i_2,l_2}\mid\cF_n}$, $\sigma^{\bZ}_{k_1,k_2,i_1,l_1,i_2,l_2} = \eE\pth{Z_{k_1,i_1,l_1}Z_{k_2,i_2,l_2}\mid\cF_n}$. From the proof of Theorem 3 of \cite{bai2025time}, we have 
    \begin{align*}
      \left\| \max_{k_1,k_2, i_1,l_1,i_2,l_2}|\sigma^{\bZ^{\dagger}}_{k_1,k_2, i_1,l_1,i_2,l_2} - \sigma^{\bZ}_{k_1,k_2, i_1,l_1,i_2,l_2}  |\right\|_{q/2} = O\left(\frac{(\log n)^2}{w} + \frac{w}{nb} + \sqrt{\frac{w}{nb}}(np^2)^{4/q}\right).
    \end{align*}
    Denote $\vartheta_n = \frac{\log^2 n}{w} + \frac{w}{nb} + \sqrt{\frac{w}{nb}}(n p^2)^{4/q}$. Using Remark 4.1 of \cite{chernozhukov2017central}, we have \begin{align}\label{pfthm2eq4}
        \sup_{\bx \in \eR^{|\cH|}}\left|P\left(\left |\mf Z^{\dagger}|_{\infty,p}\leq \bx \right|\cF_n\right) - P\left(\left| \frac{1}{\sqrt{n b}}\sum_{i=1}^{2 \nb} \bZ_i \right|_{\infty,p} \leq \bx\right)\right| = O_\eP\qth{\vartheta_n^{1/3}\sth{\log (np)}^{2/3}}.
    \end{align}
    Let $g^{\prime}_n =p^{2/q} (\sqrt{m/(n \eta^{2})}+1/m+\eta  +\sqrt{m/(nb)}(mb/n)^{-1/2q})$, $f_n = (nb)^{-1/2} b^{-1/q} p^{2/q}$, and $q_n$ be an increasing sequence such that $q_n\to \infty,q_n(f_n+g'_n)\to 0$. Define the $\cF_n$ measurable events
    \begin{align}\label{eq:AnBn}
        A_n = \left\{ \max_{i,l\in[p]} \sup_{t \in \cT} \left|{\tilde  \Gamma}^{2}_{i,l}(t) - \Gamma^{2}_{i,l}(t) \right|  > g^{\prime}_n q_n\right\}, \quad B_n = 
    \left\{ \max_{i,l\in[p]} \sup_{t \in \cT} \left|\tilde \beta_{i,l}(t) - \beta_{i,l}(t) \right|  > f_n q_n\right\}.
    \end{align}
    By Condition \ref{Ass:bdrate}, we have $n^{\phi - 1}b^{-1}h + b^2 = O\{(nb)^{-1/2}\}$. By Lemma \ref{lem:lowodker} and Proposition 2 of \cite{bai2025time}, we have $\eP(A_n\cup B_n) = O\pth{q_n^{-q}}$. Similar to Equation (46) and (47) of \cite{bai2025time}, for some sufficiently large constant $M$ we  have \begin{align}
        \eE(|\mf Z^{\diamond} - \mf Z^{\dagger}|^q_{\infty}&\mf 1(\bar A_n \cap \bar B_n) |\cF_n) \leq  M\left| \frac{\log (np)}{2w\nb}\max_{\substack{i,l\in[p],i>l\\0 \leq a\leq  n-2\nb }}\sum_{j=w}^{2\nb - w}(\hat{\tilde S}_{a,j,i,l} - \tilde S_{a,j,i,l})^2 \mf 1(\bar A_n \cap \bar B_n) \right|^{q/2},\label{pfthm2eq2}\\
        &\frac{1}{\sqrt{2w\nb}}\left\| \max_{\substack{i,l\in[p],i>l\\0 \leq a\leq  n-2\nb }}\left(\sum_{j=w}^{2\nb - w}(\hat{\tilde S}_{a,j,i,l} - \tilde S_{a,j,i,l})^2 \mf 1(\bar A_n \cap \bar B_n)\right)^{1/2}  \right\|_q \notag\\ 
        &= O((n^{\phi}/\sqrt{w} +w^{3/2}/n  +  g^{\prime}_nq_n + f_nq_n) (np^2)^{1/q}).\label{pfthm2eq3}
    \end{align}
    Recall that we have defined $g_n = g_n' + p^{2/q}(nb)^{-1/2}b^{-1/q} + w^{3/2}/n + n^{\phi} h/\sqrt{w}$. Pick $q_n = \left\{\left(g^{\prime}_n + f_n +  w^{3/2}/n +  n^{\phi} h/\sqrt{w}\right) (np^2)^{1/q}\right\}^{-1/(q+2)} = \sth{g_n(np^2)^{1/q}}^{-1/(q+2)}$. From \eqref{bdrateeq}, we know that $q_n\rightarrow\infty, (g'_n+f_n)q_n \le g_nq_n\rightarrow 0$. By \eqref{pfthm2eq4}, \eqref{pfthm2eq2}, \eqref{pfthm2eq3}, Markov inequality and Lemma \ref{lm1}, via some direct but tedious calculation, \begin{align*}
        \sup_{
        \bx \in \eR^{|\cH|}} \left| P(|\bZ_{\bt}|_{\infty,p} \leq 
        \bx| \cF_n) - P\left(\left| \frac{1}{\sqrt{n b}}\sum_{i=1}^{2 \nb -1} \mf Z_i \right|_{\infty,p} \leq \bx\right)\right|  = & O_\eP\sth{\vartheta_n^{1/3}\pth{\log n}^{2/3}  +q_n^{-q}\log n}\\
        = & O_\eP\big(\theta^{(2)}_n\big) = o_\eP(1).
   \end{align*}
    Thus we finish the proof of \eqref{pfthm2eq1}.

\subsection{Proof of Theorem \ref{thm:pvalp}}

\subsubsection{Proof of (\romannumeral 1)}

First, by the proof of Theorem \ref{thm1} and Theorem \ref{thm2}, we have \begin{align*}
        &\sup_{\bx \in \eR^{|\cH|}} \ath{P(|Z_{\bt}|_{\infty,p} \leq \bx| \cF_n) - P\left(\sup_{t \in \cT} \frac{\sqrt{nb_{i,l}}|\tilde \rho_{i,l}(t) - \rho_{i,l}(t)|}{\tilde \Gamma_{i,l}(t)}\leq x_{i,l},i,l\in[p]\right)} = O_\eP\pth{\theta_n^{(1)} + \theta_n^{(2)}}.
    \end{align*}
    Pick $\Delta_n$ such that $\Delta_n\rightarrow 0, \Delta_n/\pth{\theta_n^{(1)} + \theta_n^{(2)}}\rightarrow\infty$. Define the event  \begin{align}\label{eq:Asigman}
        A_{\Delta_n} = &\sth{\sup_{\bx \in \eR^{|\cH|}} \ath{P(|Z_{\bt}|_{\infty,p} \leq \bx| \cF_n) - P\left(\sup_{t \in \cT} \frac{\sqrt{nb_{i,l}}|\tilde \rho_{i,l}(t) - \rho_{i,l}(t)|}{\tilde \Gamma_{i,l}(t)}\leq x_{i,l},i,l\in[p]\right)} \le \Delta_n}.
    \end{align}
    We have $\eP(\bar{A}_{\Delta_n})\rightarrow 0$. Using Lemma \ref{lem:pvalE}(\romannumeral 1) and letting $n,B\rightarrow\infty$, we finish the proof of Theorem \ref{thm:pvalp}(\romannumeral 1).

    \subsubsection{Proof of (\romannumeral 2)}

    Define $\tilde{\rho}_{i,l}^*,\tilde{\Gamma}_{i,l}^*$ as an i.i.d. copy of $\tilde{\rho}_{i,l},\tilde{\Gamma}_{i,l}$. Let $x$ be a fixed number in $(0,1)$. Define $\Omega^*(\tilde{\rho}_{i,l},\tilde{\rho}_{i,l}^*) : = \sth{\frac{\sqrt{nb_{i,l}}|\tilde \rho_{i,l}(t)|}{\tilde \Gamma_{i,l}(t)} < \sup_{t\in\cT}\frac{\sqrt{nb_{i,l}}|\tilde \rho_{i,l}^*(t) - \rho_{i,l}(t)|}{\tilde \Gamma_{i,l}^*(t)}}$. Following the procedure in \eqref{pflmB5eq1}, we have \begin{align}
        &\eP\sth{P_{i,l}(t) \ge x,A_{\Delta_n}} = \eE\qth{\eP\sth{P_{i,l}(t) \ge x\mid\cF_n}\mathbf{1}(A_{\Delta_n})}\notag\\
        \le & \eE\qth{\eP\sth{\mathrm{Binom}\pth{\qth{\eP\sth{T_{i,l}(t) < \sup_{t \in \cT} \frac{\sqrt{nb_{i,l}}|\tilde \rho_{i,l}^*(t) - \rho_{i,l}(t)|}{\tilde \Gamma_{i,l}^*(t)} \mid\cF_n}+\Delta_n} \vee 0,B}\ge \lceil xB\rceil}}\notag\\
        = & \eE\qth{\eP\sth{\mathrm{Binom}\pth{\qth{\eP\sth{\Omega^*(\tilde{\rho}_{i,l},\tilde{\rho}_{i,l}^*) \mid\cF_n}+\Delta_n} \vee 0,B}\ge \lceil xB\rceil}}.\label{eq:Binomcompare}
    \end{align}
    To further evaluate $\eP\sth{\Omega^*(\tilde{\rho}_{i,l},\tilde{\rho}_{i,l}^*) \mid\cF_n}$, we define the following three events: 
    for positive constants $M,M'$, define $D_M^* = \sth{\sup_{t\in\cT}\frac{\sqrt{nb_{i,l}}|\tilde \rho_{i,l}^*(t) - \rho_{i,l}(t)|}{\tilde \Gamma_{i,l}^*(t)} > M}$, $D_M = \Big\{\frac{\sqrt{nb_{i,l}}|\tilde \rho_{i,l}(t) - \rho_{i,l}(t)|}{\tilde \Gamma_{i,l}(t)} > M\Big\},$ and $E_{M'}= \sth{\tilde \Gamma_{i,l}(t) > M'}~$. By Lemma \ref{lem:lowodker}, \eqref{beq01}, and Proposition 2 of \cite{bai2025time}, we have $\sup_{t\in\cT} \sqrt{nb_{i,l}}|\tilde \rho_{i,l}(t) - \rho_{i,l}(t)|/\tilde \Gamma_{i,l}(t) = O_\eP(1), 1/\tilde \Gamma_{i,l}(t) = O_\eP(1), \tilde \Gamma_{i,l}(t) = O_\eP(1)$, which gives  $\underset{M,M'\rightarrow\infty}{\lim}\underset{n\rightarrow\infty}{\limsup}~\eP(D_M\cup E_{M'}) = 0, ~\underset{M\rightarrow\infty}{\lim}\underset{n\rightarrow\infty}{\limsup}~\eP(D_M^*) = 0$.
    We have the following decomposition: \begin{align*}
        &\eP\sth{\frac{\sqrt{nb_{i,l}}|\tilde \rho_{i,l}(t)|}{\tilde \Gamma_{i,l}(t)}\le \sup_{t\in\cT}\frac{\sqrt{nb_{i,l}}|\tilde \rho_{i,l}^*(t) - \rho_{i,l}(t)|}{\tilde \Gamma_{i,l}^*(t)} \mid\cF_n} = I_n^{(1)} + I_n^{(2)}, ~~\mbox{where} \\
        & I_n^{(1)}= \eP\sth{\frac{\sqrt{nb_{i,l}}|\tilde \rho_{i,l}(t)|}{\tilde \Gamma_{i,l}(t)}\le \sup_{t\in\cT}\frac{\sqrt{nb_{i,l}}|\tilde \rho_{i,l}^*(t) - \rho_{i,l}(t)|}{\tilde \Gamma_{i,l}^*(t)}, D_M^* \mid\cF_n},\\
        &I_n^{(2)} = \eP\sth{\frac{\sqrt{nb_{i,l}}|\tilde \rho_{i,l}(t)|}{\tilde \Gamma_{i,l}(t)}\le \sup_{t\in\cT}\frac{\sqrt{nb_{i,l}}|\tilde \rho_{i,l}^*(t) - \rho_{i,l}(t)|}{\tilde \Gamma_{i,l}^*(t)}, \overline{D_M^*} \mid\cF_n}.
    \end{align*}
    We first evaluate $I_n^{(1)}$. Notice that $\tilde{\rho}_{i,l}^*(t),\tilde{\Gamma}_{i,l}^*(t)$ are an i.i.d. copy of $\tilde{\rho}_{i,l}(t),\tilde{\Gamma}(t)$, so $D_M^*$ is independent of $\cF_n$. This gives $ I_n^{(1)} \le \eP(D_M^*\mid \cF_n) = \eP(D_M^*),$ and $ ~\lim_{M\rightarrow\infty}\limsup_{n\rightarrow\infty} I_n^{(1)} \rightarrow_\eP 0$. 
    
    Next we evaluate $I_n^{(2)}$. We have $I_n^{(2)} \le \eP\sth{\frac{\sqrt{nb_{i,l}}|\tilde \rho_{i,l}(t)|}{\tilde \Gamma_{i,l}(t)}\le M \mid\cF_n}$. Noticing that $\sqrt{nb_{i,l}}|\tilde \rho_{i,l}(t)|/\tilde \Gamma_{i,l}(t)$ is $\cF_n$-measurable, we have \begin{align*}
        I_n^{(2)} \le & \mathbf{1}\sth{\frac{\sqrt{nb_{i,l}}|\tilde \rho_{i,l}(t)|}{\tilde \Gamma_{i,l}(t)}\le M, \overline{D_M}\cap \overline{E_{M'}}} + \mathbf{1}(D_M\cup E_{M'})\\ 
        \le & \mathbf{1}\sth{\frac{\sqrt{nb_{i,l}}| \rho_{i,l}(t) | - \sqrt{nb_{i,l}}|\tilde \rho_{i,l}(t) - \rho_{i,l}(t)|}{\tilde \Gamma_{i,l}(t)}\le M, \overline{D_M}\cap \overline{E_{M'}}} + \mathbf{1}(D_M\cup E_{M'})\\
        \le & \mathbf{1}\sth{\sqrt{nb_{i,l}}| \rho_{i,l}(t)| \le 2M M'} + \mathbf{1}(D_M\cup E_{M'}).
    \end{align*}
    For any fixed $M,M'$, let $n\rightarrow \infty$, we have $\sqrt{nb_{i,l}}|\rho_{i,l}(t)|\rightarrow\infty$, so $\limsup_{n\rightarrow\infty} I_{n}^{(2)} \le \limsup_{n\rightarrow\infty} I(D_M\cup E_{M'})$.
    So letting $M,M'\rightarrow\infty$, we have $I_n^{(2)} \rightarrow_\eP 0$.
    
    Define event $\Omega_{i,l,n}(x) = \sth{\eP\sth{\Omega^*(\tilde{\rho}_{i,l},\tilde{\rho}_{i,l}^*) \mid\cF_n} < x/2}$. 
    By the analysis of $I_n^{(1)}, I_n^{(2)}$, we have $\eP\sth{\Omega^*(\tilde{\rho}_{i,l},\tilde{\rho}_{i,l}^*) \mid\cF_n} \rightarrow_\eP 0$, and thus $\eP\sth{\overline{\Omega_{i,l,n}(x)}} \rightarrow 0$. By \eqref{eq:Binomcompare}, we have \begin{align*}
        &\eP\sth{P_{i,l}(t) \ge x,A_{\Delta_n}}\\
        \le & \eE\qth{\eP\sth{\mathrm{Binom}\pth{\qth{\eP\sth{\Omega^*(\tilde{\rho}_{i,l},\tilde{\rho}_{i,l}^*)\mid\cF_n}-\Delta_n} \vee 0,B}\ge \lceil xB\rceil}}\\
        \le & \eE\qth{\mathbf{1}\sth{\Omega_{i,l,n}(x)}\eP\sth{\mathrm{Binom}\pth{\qth{\eP\sth{\Omega^*(\tilde{\rho}_{i,l},\tilde{\rho}_{i,l}^*)\mid\cF_n}-\Delta_n} \vee 0,B}\ge \lceil xB\rceil}}  +\eP\sth{\overline{\Omega_{i,l,n}(x)}}\\
        \le & \eP\qth{\mathrm{Binom}\sth{\pth{x/2+\Delta_n} \wedge 1,B}\ge \lceil xB\rceil} +\eP\sth{\overline{\Omega_{i,l,n}(x)}} \rightarrow 0.
    \end{align*}
    In the last line we use the fact that $B\rightarrow\infty, \Delta_n \rightarrow 0$. Since $\Delta_n/\{\theta_n^{(1)}+\theta_n^{(2)}\}\rightarrow\infty$, we have $\eP(\bar{A}_{\Delta_n})\rightarrow 0$ and we finish the proof.


    \subsection{Proof of Theorem \ref{thm3}}

    Recall the definition of $A_{\Delta_n}$ in \eqref{eq:Asigman}. Choose $\Delta_n$ such that $|\cH|\Delta_n\rightarrow 0$ and $\{\theta_n^{(1)}+\theta_n^{(2)}\}/\Delta_n\rightarrow 0$; such a sequence exists because $p^2\{\theta_n^{(1)}+\theta_n^{(2)}\}\rightarrow 0$. By the approximation bounds in Theorems \ref{thm1}--\ref{thm2}, this choice gives $\eP(\bar{A}_{\Delta_n})\rightarrow 0$.

    Write $\cH_0^\star=\cup_{t\in[0,1]}\cH_0(t)$, so that $|\cH_0^\star|/|\cH|=\pi_0^\star$. Let
    \[
        \mathcal{B}_n=([0,1]\backslash\cT)\cup \bigcup_{(i,l)\in\cH}(\cI_{i,l}\backslash \cI_{i,l}')
    \]
    be the time region affected by boundary effects or by changes in the true-null set. We have $\mathrm{Leb}(\mathcal{B}_n)\le 2|\cC_0|b+2b$. Therefore,
    \begin{align*}
        \mathrm{AuFDR}_{r_n}
        = \eE\{ \|\fdp(\cdot)\|_{r_n}\}
        &\le \{\mathrm{Leb}(\mathcal{B}_n)\}^{1/r_n}
        + \eE\sth{\sup_{t\in[0,1]\backslash \mathcal{B}_n}\fdp(t)}.
    \end{align*}
    The first term converges to zero by $|\cC_0|b=O(n^{-\iota})$ for some $\iota >0$, and $r_n=o(\log n)$.
    
    It remains to bound the second term. For $t\notin\mathcal{B}_n$ and $(i,l)\in\cH_0(t)$, we have $t\in\cI_{i,l}'$. Under the time-varying BY procedure, if $(i,l)\in\cR(t)$, then
    \[
        P_{i,l}(t)\le \frac{\alpha |\cR(t)|}{|\cH|\ell(|\cH|)}.
    \]
    Hence $P_{i,l}(t)\le \alpha/\ell(|\cH|)$ and $|\cR(t)|\ge \lceil |\cH|\ell(|\cH|) P_{i,l}(t)/\alpha\rceil$. Consequently,
    \begin{align*}
        \sup_{t\in[0,1]\backslash \mathcal{B}_n}\fdp(t)
        &\le
        \sum_{(i,l)\in\cH_0^\star}
        \sup_{t\in\cI_{i,l}'}
        \frac{I\{P_{i,l}(t)\le \alpha/\ell(|\cH|)\}}
        {\lceil |\cH|\ell(|\cH|) P_{i,l}(t)/\alpha\rceil\vee 1}.
    \end{align*}
    For each $(i,l)\in\cH_0^\star$, let $P_{i,l,\min}=\inf_{t\in\cI_{i,l}'}P_{i,l}(t)$; if $\cI_{i,l}'=\emptyset$, the corresponding summand is set to zero. Since the function $x\mapsto I(x\le \alpha/\ell(|\cH|))/\{\lceil |\cH|\ell(|\cH|)x/\alpha\rceil\vee 1\}$ is nonincreasing, we have
    \begin{align*}
        \sup_{t\in[0,1]\backslash \mathcal{B}_n}\fdp(t)
        \le
        \sum_{(i,l)\in\cH_0^\star}
        \frac{I(P_{i,l,\min}\le \alpha/\ell(|\cH|))}
        {\lceil |\cH|\ell(|\cH|){P_{i,l,\min}}/\alpha\rceil\vee 1}.
    \end{align*}
    
    We now bound the expectation of each summand on $A_{\Delta_n}$. By Lemma \ref{lem:pvalE}, for every $x\in[0,1]$,
    \[
        \eP(P_{i,l,\min}\le x\mid A_{\Delta_n})\le x+\Delta_n+B^{-1}.
    \]
    Let $t_k=k\alpha/\{|\cH|\ell(|\cH|)\}$ for $k=0,\ldots,|\cH|$, and denote $F_k=\eP(P_{i,l,\min}\le t_k\mid A_{\Delta_n})$ with $F_0=0$ in the following summation-by-parts bound. Then
    \begin{align*}
        \eE\sth{
        \frac{I\{P_{i,l,\min}\le \alpha/\ell(|\cH|)\}}
        {\lceil |\cH|\ell(|\cH|) P_{i,l,\min}/\alpha\rceil\vee 1}
        \,\bigg|\, A_{\Delta_n}}
        &\le \sum_{k=1}^{|\cH|} \frac{F_k-F_{k-1}}{k}  \\
        &= \frac{F_{|\cH|}}{|\cH|}+\sum_{k=1}^{|\cH|-1}F_k\pth{\frac1k-\frac{1}{k+1}}\\
        &\le \frac{\alpha}{|\cH|}+\Delta_n+B^{-1}.
    \end{align*}
    The last inequality follows from the fact that, on $A_{\Delta_n}$, $F_k\le k\alpha/(|\cH|\ell(|\cH|))+\Delta_n+B^{-1}$ and $\sum_{k=1}^{|\cH|} k^{-1}=\ell(|\cH|)$.

    Since $\fdp(t)\le 1$, we obtain
    \begin{align*}
        \eE\sth{\sup_{t\in[0,1]\backslash \mathcal{B}_n}\fdp(t)}
        &\le \eP(\bar{A}_{\Delta_n})
        + |\cH_0^\star|\sth{\frac{\alpha}{|\cH|}+\Delta_n+B^{-1}}\\
        &\le \alpha \pi_0^\star+\eP(\bar{A}_{\Delta_n})
        +|\cH|\Delta_n+|\cH|B^{-1}.
    \end{align*}
    By the choice of $\Delta_n$ and the assumption $Bp^{-2}\rightarrow\infty$, the last three terms converge to zero. Therefore,
    \[
        \limsup_{n\to\infty}(\mathrm{AuFDR}_{r_n}-\alpha\pi_0^\star)\le 0,
    \]
    which proves Theorem \ref{thm3}.
    
    \section{Additional Technical Proofs}\label{app:SecE}
    
    \subsection{Lemma \ref{lm1} and its Proof}

\begin{lemma}\label{lm1}
    For any $n|\cH|$-dimensional random vectors $\bX,\bX'$ and $n|\cH|$-dimensional Gaussian random vectors $\bY$, assume there exists a positive constant $C_1$ such that $\eE(Y_j^2) > C_1$ for $j\in[n|\cH|]$. Recall the definition of norm $|\cdot|_{\infty,p}$ in \eqref{eq6}. Then for every $\by\in\eR^{|\cH|}$ and $\delta > 0$, \begin{align*}
        \sup_{\by\in\eR^{|\cH|}}\ath{\eP(|\bX|_{\infty,p} < \by)-\eP(|\bY|_{\infty,p} < \by)} \le & \sup_{\by\in\eR^{|\cH|}}\ath{\eP(|\bX'|_{\infty,p} < \by)-\eP(|\bY|_{\infty,p} < \by)}\\
        &+ \eP(|\bX-\bX'|_{\infty}>\delta)+ C\delta\sqrt{\log \pth{n|\cH|}},
    \end{align*}
    where $C$ is a positive constant that only depends on $C_1$.
\end{lemma}

\textbf{Proof.} Lemma \ref{lm1} can be proved in a similar way as the proof of Lemma S1 in \cite{dette2021confidence}. For simplicity we omit the proof.

\subsection{Lemma \ref{lm3} and its Proof}

\begin{lemma}\label{lm3}
    Assume all conditions of Theorem \ref{thm1}.  Define \begin{align*}
        H_{i,l,h}(t,\cF_j) &= \sth{G_i(t,\cF_j)-G_i(t,\cF_{j-h})}\sth{G_l(t,\cF_j)-G_l(t,\cF_{j-h})},\\
        H_{i,l,\bXi}(t,\cF_j) &= \frac{H_{i,l,h}(t,\cF_j)}{2 \sigma_{i,l}(t)} - \frac{\rho_{i,l}(t)}{4}\sth{\frac{H_{i,i,h}(t,\cF_j)}{\gamma_{i,i}(t)}+\frac{H_{l,l,h}(t,\cF_j)}{\gamma_{l,l}(t)}}.
    \end{align*}
    Then we have (\romannumeral 1) $H_{i,l,\bXi}\in\mathrm{Lip}_{q}$ and their Lipschitz constant are uniformly bounded over $i,l\in[p]$. (\romannumeral 2) There exists a positive constant $t_1$ such that $\sup_{t\in[0,1],i,l\in[p]}\eE\sth{\exp\pth{t_1|H_{i,l,\bXi}(t,\cF_j)|^\kappa}} < \infty$. (\romannumeral 3) $\max_{i,l\in[p]}\delta_{q}(H_{i,l,\bXi},r) = O(\chi^{r-h})$ for some $\chi\in(0,1)$.
\end{lemma}

    \textbf{Proof.} Notice that we have assumed $G_i(t,\cF_j)\in\mathrm{Lip}_{2q}, \sup_{t\in[0,1],i\in[p]}\nth{G_i(t,\cF_j)}_{2q} < \infty$ in (\romannumeral 1) of Assumption \ref{Ass:error} with uniformly bounded Lipschitz constant. Using Cauchy-Schwarz inequality, $H_{i,l,h}(t,\cF_j)\in\mathrm{Lip}_{q}$ and their Lipschitz constants are also uniformly bounded. Then (\romannumeral 1) comes from the above claim and the fact that $\gamma_{i,i}(t),\gamma_{l,l}(t),\sigma_{i,l}(t)$ are all lower and upper bounded Lipschitz continuous functions with bounded Lipschitz constants as assumed in Assumption \ref{Ass:error}(\romannumeral 4), and $\rho_{i,l}(t)$ are also bounded Lipschitz continuous functions with bounded Lipschitz constants. We can prove (\romannumeral 2) using similar arguments.


    For (\romannumeral 3), using Minkowski inequality, we have \begin{align*}
        \delta_q(H_{i,l,h},r) \le & \sup_{t\in[0,1]}\nth{G_{i}(t,\cF_r)G_{l}(t,\cF_r)-G_{i}(t,\cF_r^*)G_{l}(t,\cF_r^*)}_q \\
        &+ \sup_{t\in[0,1]}\nth{G_{i}(t,\cF_r)G_{l}(t,\cF_{r-h})-G_{i}(t,\cF_r^*)G_{l}(t,\cF_{r-h}^*)}_q\\
        & + \sup_{t\in[0,1]}\nth{G_{i}(t,\cF_{r-h})G_{l}(t,\cF_r)-G_{i}(t,\cF_{r-h}^*)G_{l}(t,\cF_r^*)}_q \\
        &+ \sup_{t\in[0,1]}\nth{G_{i}(t,\cF_{r-h})G_{l}(t,\cF_{r-h})-G_{i}(t,\cF_{r-h}^*)G_{l}(t,\cF_{r-h}^*)}_q\\
        := & I_1 + I_2 + I_3 + I_4.
    \end{align*}
    For $I_1$, using Cauchy-Schwarz inequality we have \begin{align*}
        I_1 \le \sup_{t\in[0,1],i\in[p]}\nth{G_i(t,\cF_r)}_{2q}\sth{\delta_{2q}(G_i,r)+\delta_{2q}(G_l,r)}.
    \end{align*}
    Similarly bounds can be obtained for $I_2, I_3$ and $I_4$. Combining the evaluation above, we have $\delta(H_{i,l,h},r) = O(\chi^r + \chi^{r-h})$. Since $\sigma_{i,l}^{-1}(t),\rho_{i,l}(t)/\gamma_{i,i}(t),\rho_{i,l}(t)/\gamma_{l,l}(t)$ are all bounded functions, using Minkowski inequality again we can obtain (\romannumeral 3).

\subsection{Lemma \ref{lm2} and its Proof}

\begin{lemma}\label{lm2}
    Recall the definition of $\bar{\mf \Xi}^{\cG}_j$ before Theorem \ref{thm1}, and the definition of $\ath{\cdot}_{\infty,p}$ in \eqref{eq6}. Under all conditions of Theorem \ref{thm1}, there exists a sequence of zero-mean Gaussian vectors $(\mf Z_j)_{j=1}^{2\nb} = (\mf Z_{j,2,1},\dots,\mf Z_{j,p,p-1})_{j=1}^{2\nb} \in \eR^{(n - 2\nb+1)|\cH|}$, which share the same autocovariance structure with the vectors $(\bar{\mf \Xi}^{\cG}_j)_{j=1}^{2\nb} $  such that \begin{align*}
    \sup_{\bx\in\eR^{|\cH|}} \ath{\eP\pth{\ath{\frac{1}{\sqrt{nb}}\sum_{j=1}^{2\lceil nb \rceil}\bar{\mf \Xi}^{\cG}_j}_{\infty,p}\le \bx} - \eP\pth{\ath{\frac{1}{\sqrt{nb}}\sum_{j=1}^{2\lceil nb \rceil}\mf Z_j}_{\infty,p}\le \bx}} = O\sth{(nb)^{-1/8}\pth{\log n}^{4}}.
\end{align*}
\end{lemma}

    \textbf{Proof.} Recall the definition of $\Xi_{j,i,l}$ in \eqref{eq:Xidiscrete}, the definition of $\Xi_{j,s,i,l}$ in \eqref{eq:defXjsz}, and the definition of $\Bar\bXi_{j,i,l}$ before Theorem \ref{thm1}. Define $\Bar{\bXi}_{j,s}^\cG,j,s\in[n] = (\Xi_{j,s,2,1},\Xi_{j,s,3,1},\Xi_{j,s,3,2},\dots,\Xi_{j,s,p,p-1})^\T$.  Let $\bXi_{j}^{\cG} = (\Xi_{j,2,1},\dots,\Xi_{j,p,p-1})$, and $\bXi_{j}^{\cG,W} = (\Bar{\bXi}_{j,1}^\cG,\dots,\Bar{\bXi}_{j,n}^\cG)$. Define the $M$-dependent version of $\Xi_{j,i,l}$ as $\Xi_{j,i,l}^{(M)} = \eE\pth{\Xi_{j,i,l}\mid \xi_{j-M+1},\dots,\xi_j}$, and the corresponding $M$-dependent version of $\bXi_{j}^{\cG},\bXi_{j}^{\cG,W}$ are denoted as $\bXi_{j}^{\cG,(M)},\bXi_{j}^{\cG,W,(M)}$. Define $\bZ_j^{\cG},\bZ_j^{\cG,W},\bZ_j^{\cG,W,(M)}$ as the normal vectors with the same covariance structure as $\bXi_j^{\cG},\bXi_j^{\cG,W},\bXi_j^{\cG,W,(M)}$. Define $\bXi^{\cG,W} = (nb)^{-1/2}\sum_{j=1}^n \bXi^{\cG,W}_j$ and similarly define $\bXi^{\cG,W,(M)},\bZ^{\cG,W},\bZ^{\cG,W,(M)}$.
    
    Define $A\in\cA^{Re}$ if  $A=\sth{\bw = (w_1,\dots,w_{n|\cH|})\in\eR^{n|\cH|}:a_j\le w_j\le b_j,j\in[n|\cH|]}$, and $-\infty \le a_j,b_j\le\infty$. Let $\beta = \varphi\log (n|\cH|)$, where $\varphi$ is a rescaling constant. Define $F_\beta(\bw) = \beta^{-1}\log\sth{\sum_{i=1}^{n|\cH|}\exp\pth{\beta w_i}},\bw=(w_1,\dots,w_{n|\cH|})\in\eR^{n|\cH|}$. Let $g_0:\eR\rightarrow [0,1]$ be a continuously differentiable function with finite third order derivatives and $g(u) = g_0(\varphi u)$. Define $m(\bw) = g\sth{F_\beta(\bw)},\bw\in\eR^{n|\cH|}$ and $G_k=\sup_{x\in\eR}|\partial^kg(x)/\partial x^k|$ for $k\ge 0$. Define $M_x,M_y$ as a truncation level for $\bXi^\cG_j,\bZ_j$, which appears in the intermediate step of the proof of Proposition 6 in \cite{wu2024asynchronous} (detailed in the definition of $\tilde{x}_{ij},\tilde{y}_{ij}$ in their paper). For bandwidth $b_{i,l}$, define $b = \max_{i,l} b_{i,l},b' = \min_{i,l} b_{i,l}$.
    
    Results in Lemma \ref{lm3} about $H_{i,l,\bXi}$ corresponds to Condition (A1)-(A3) of \cite{wu2024asynchronous}. Define $b'=\min_{i,l} b_{i,l}$. With a very careful investigation of the proof of Proposition 6 and 7 of \cite{wu2024asynchronous}, we have equation (236) still holds in view of results in Lemma \ref{lm3} and Assumption \ref{Ass:error} with minor modifications, which is, for some small enough but positive constant $t$, \begin{align}
        &|\eE\sth{m(\bXi^{\cG,W}) - m(\bZ^{\cG,W,(M)})}| \lesssim (\varphi^2+\varphi\beta)(M_x^{-1}+M_y^{-1}) + (\varphi^3+\varphi^2\beta+\varphi\beta^2)M^2(nb')^{-1/2}\notag\\
        &  + (\varphi^qn|\cH|\chi^{(M-h)q})^{1/(1+q)} + \frac{\sqrt{nb\log n}}{M_x^q} + n|\cH|\sth{\exp(-tM_x^{\kappa})+\exp(-tM_y^2)}.\label{pflm2eq2}
    \end{align}
    subject to $\sqrt{5}(4M+1)(M_x\vee M_y)/\sqrt{nb'}\le \beta^{-1}$. 
    
    The first difference comes from the term $(\varphi^qn|\cH|\chi^{(M-h)q})^{1/(1+q)}$, which results from the fact that in (\romannumeral 3) of Lemma \ref{lm3} we derive exponential decay of $\max_{i,l\in[p]}\delta_q(H_{i,l,\bXi},l) = O(\chi^{l-h})$, while in Condition (A3) of \cite{wu2024asynchronous} they assume $\max_{i,l\in[p]}\delta_1(H_{i,l,\bXi},l)=O(\chi^l)$. The second difference is the term $\exp(-tM_x^{\kappa})$. This is due to the tail results of $\sup_{t\in[0,1],i,l\in[p]}\eE\sth{\exp\pth{t_0|H_{i,l,\bXi}(t,\cF_j)|^\kappa}} < \infty$ in Lemma \ref{lm3}, which is different from exponential tail condition $\sup_{t\in[0,1],i,l\in[p]}\eE\sth{\exp\pth{t_0|H_{i,l,\bXi}(t,\cF_j)|}} < \infty$ as imposed in \cite{wu2024asynchronous}. Recall $\vartheta_{i,l}(t)=(nb_{i,l})^{-1}\sum_{j=1}^n K_{b_{i,l}}(t_j-t)\bXi_{j,i,l}$. By elementary calculation similar to Lemma C.3 in \cite{dette2021confidence}, noticing that $h\asymp \log n$, we have uniformly for $i,l\in[p]$, \begin{align}
        nb_{i,l}\eE\sth{\vartheta_{i,l}^2(t)} = \tilde{\Gamma}_{i,l}^2(t) + O\sth{b\log b +\chi^{nbr_n-h} + r_n + (nb)^{-1}}. \label{beq01}
    \end{align}
    Here $r_n = (M'\log n+h)/(nb) \le M''\log n/(nb)$ and $M', M''$ are two large enough constants. From Assumption \ref{Ass:error}(\romannumeral 5), using argument in Lemma E.4 of \cite{wu2024multiscale}, we have for large enough $n$, there exists positive constant $c_1,c_2$ such that $c_1 < \var(\bXi_{i,l}^{\cG,W})<c_2 ,i,l\in[p]$, with $\bXi_{i,l}^{\cG,W}$ being $z$-th element of $\bXi^{\cG,W}$. Using argument similar to Lemma A.2 of \cite{zhang2018gaussian}, noticing that $\sum_{l=M+1}^\infty l\sup_{i,l\in[p]}\delta_q(H_{i,l,\Xi},l)\rightarrow 0$ if $M = a \log n$ and $a$ is a large enough constant, we have there exist positive constants $c_3,c_4$ such that $c_3 < \var(\bXi_{i,l}^{\cG,W}), \var(\bXi_{i,l}^{\cG,W,(M)})< c_4 ,i,l\in[p]$. Using argument of step 5 of \cite{zhang2018gaussian} we can derive $\ath{\var\sth{\bXi^{\cG,W}}-\var\sth{\bXi^{\cG,W,(M)}}}_{\max}\lesssim \max_{i,l\in[p]}\sum_{l=M+1}^\infty l\delta_q(H_{\bXi,z},l)\lesssim (\chi')^{M-h}$, where $\chi'$ is another constant satisfying $\chi'\in(\chi,1)$.
    Noticing that $\bZ^{\cG,W},\bZ^{\cG,W,(M)}$ has the same covariance structure as $\bXi^{\cG,W},\bXi^{\cG,W,(M)}$, by Proposition 8 of \cite{wu2024asynchronous}, we have \begin{align}\label{pflm2eq1}
        \sup_{\bx\in\eR^{n|\cH|}}|\eP(\bZ^{\cG,W}\le \bx)-\eP(\bZ^{\cG,W,(M)}\le \bx)|\lesssim (\chi')^{(M-h)/3}\log^{2/3}(n).
    \end{align}
    By \eqref{pflm2eq2}, \eqref{pflm2eq1} and tedious calculation the same as the end of proof of Proposition 7 in \cite{wu2024asynchronous}, we have \begin{align*}
        \sup_{A\in\cA^{Re}}\ath{\eP\pth{\bXi^{\cG,W}\in A} - \eP\pth{\bZ^{\cG,W}\in A}} = O\sth{\frac{\pth{\log n}^{4}}{(nb)^{1/8}}}.
    \end{align*}
    And finally, Lemma \ref{lm2} comes from the above equation and the fact $
        \ath{\frac{1}{\sqrt{nb}}\sum_{j=1}^n\bXi_j^{\cG,W}}_{\infty,p} = \ath{\frac{1}{\sqrt{nb}}\sum_{j=1}^{2\lceil nb \rceil}\bar{\bXi}_j^{\cG}}_{\infty,p},\quad\ath{\frac{1}{\sqrt{nb}}\sum_{j=1}^n \bZ_j^{\cG,W}}_{\infty,p} = \ath{\frac{1}{\sqrt{nb}}\sum_{j=1}^{2\lceil nb \rceil}\bZ_j^{\cG}}_{\infty,p}$.

\subsection{Lemma \ref{lem:lowodker} and its Proof}

\begin{lemma}\label{lem:lowodker}
    Assume that all assumptions in Theorem \ref{thm1}  and \ref{thm2} hold. Then we have:

    (\romannumeral 1) The difference-based estimator has the following approximation, namely \begin{align*}
        \max_{i,l} \nth{\sup_{t\in\cT}\ath{\widehat{\beta}_{i,l}(t)-\beta_{i,l}(t) - \frac{1}{nb_{i,l}}\sum_{j=1}^n K_{b_{i,l}}(t_j-t) e_{j,i,l}}}_q = O\{b^{-1/q} (n^{\phi - 1}b^{-1}h + b^2)\}.
    \end{align*}

    (\romannumeral 2) The time-varying correlation estimator has the following approximation, namely \begin{align*}
        \max_{i,l} \nth{\sup_{t\in\cT}\ath{\tilde{\rho}_{i,l}(t)-\rho_{i,l}(t) - \frac{1}{nb_{i,l}}\sum_{j=1}^n K_{b_{i,l}}(t_j-t) \Xi_{j,i,l}}}_q = O\{b^{-1/q} (n^{\phi - 1}b^{-1}h + b^2)\}.
    \end{align*}
\end{lemma}

    \textbf{Proof.} Lemma \ref{lem:lowodker} follows from identical analysis in the proof of Lemma 1 and Proposition 1 of \cite{bai2025time}. The only difference originates from the fact that we do not assume $\int_{-\infty}^{\infty} u^2 K(u) du = 0$, and $n^{-1/2}h = O(b^2)$.

\subsection{Lemma \ref{lem:plugingap} and its Proof.}

\begin{lemma}\label{lem:plugingap}
    Assume all conditions in Theorem \ref{thm1} hold. Then we have \begin{align*}
        &\eP\sth{\max_{i,l\in[p]}\ath{\sup_{t\in\cT}\ath{\frac{\sqrt{nb_{i,l}}|\tilde \rho_{i,l}(t) - \rho_{i,l}(t)|}{ \Gamma_{i,l}(t)}}-\sup_{t\in\cT}\ath{\frac{\sqrt{nb_{i,l}}|\tilde \rho_{i,l}(t) - \rho_{i,l}(t)|}{\tilde \Gamma_{i,l}(t)}}}>\delta} \\
        &\le \delta^{-q/2}\sth{p^2b^{-1/2}(g^{\prime}_n)^{q/2}} + O\qth{(\sqrt{nb}c_n)^{q/(q+1)} + \sth{g_n(np^2)^{1/q}}^{q/(q+2)}}.
    \end{align*}
    Here $b = \max_{i,l} b_{i,l},\cT=[b,1-b]$, $c_n$ is defined in \eqref{pfthm1eq2}, $g^{\prime}_n =p^{2/q} (\sqrt{m/(n \eta^{2})}+1/m+\eta  +\sqrt{m/(nb)}(mb/n)^{-1/2q})$, $g_n = g^{\prime}_n + p^{2/q} (nb)^{-1/2}b^{-1/q} + w^{3/2}/n + n^{\phi} h/\sqrt{w}$ and $q_n = \sth{g_n(np^2)^{1/q}}^{-1/(q+2)}$, $f_n=(nb)^{-1/2}(p^2/b)^{1/q}$.
\end{lemma}

    \textbf{Proof.} Define event $A_n$ as in \eqref{eq:AnBn}, and $B_n' = \sth{\max_{i,l\in[p]}\sup_{t\in\cT}\ath{\tilde{\gamma}_{i,l}(t)-\gamma_{i,l}(t)} > f_nq_n'}$, where $q_n'$ is a positive sequence such that $q_n'\rightarrow\infty$ and $f_nq_n'\rightarrow 0$. Recall that in the proof of Theorem \ref{thm1}, we set $q'_n = (\sqrt{nb}c_n)^{-1/(q+1)}$. By Lemma \ref{lem:lowodker} and $\tilde\gamma_{i,l}(t) = \widehat{\beta}_{i,l}(t)/2, \gamma_{i,l}(t) = \beta_{i,l}(t)/2$, we have $\eP(B_n') = O\sth{(q^{\prime}_n)^{-q}}$. By Proposition 2 of \cite{bai2025time}, we have $\eP(A_n) = O(q_n^{-q})$. Since $q_n = \sth{g_n(np^2)^{1/q}}^{-1/(q+2)}$, from the bandwidth condition in Assumption \ref{Ass:bdrate}, we have $q_n\rightarrow\infty, g_n'q_n\le g_nq_n\rightarrow 0$.

    In Assumption \ref{Ass:error} we assume that $\inf_{t\in[0,1]}\min_{i,l\in[p]}\Gamma_{i,l}(t) > C$, where $C$ is a positive absolute constant. So when $n$ is large enough, on event $A_n$, $\inf_{t\in\cT}\min_{i,l\in[p]}\tilde{\Gamma}_{i,l}(t) > C/2$, and \begin{align*}
        &\ath{\frac{\sqrt{nb_{i,l}}|\tilde \rho_{i,l}(t) - \rho_{i,l}(t)|}{\Gamma_{i,l}(t)}-\frac{\sqrt{nb_{i,l}}|\tilde \rho_{i,l}(t) - \rho_{i,l}(t)|}{\tilde \Gamma_{i,l}(t)}} \le C \sqrt{nb_{i,l}}|\tilde \rho_{i,l}(t) - \rho_{i,l}(t)||\tilde \Gamma_{i,l}(t)-\Gamma_{i,l}(t)|,
    \end{align*}
    where $C$ is a positive absolute constant. This gives \begin{align}
        &\eP\sth{\max_{i,l\in[p]}\ath{\sup_{t\in\cT}\ath{\frac{\sqrt{nb_{i,l}}|\tilde \rho_{i,l}(t) - \rho_{i,l}(t)|}{ \Gamma_{i,l}(t)}}-\sup_{t\in\cT}\ath{\frac{\sqrt{nb_{i,l}}|\tilde \rho_{i,l}(t) - \rho_{i,l}(t)|}{\tilde \Gamma_{i,l}(t)}}}>\delta}\notag \\
        \le &\eP\sth{\max_{i,l\in[p]}\sup_{t\in\cT}\ath{\frac{\sqrt{nb_{i,l}}|\tilde \rho_{i,l}(t) - \rho_{i,l}(t)|}{ \Gamma_{i,l}(t)}-\frac{\sqrt{nb_{i,l}}|\tilde \rho_{i,l}(t) - \rho_{i,l}(t)|}{\tilde \Gamma_{i,l}(t)}}>\delta, \overline{A_n}, \overline{B_n'}} + \eP(A_n) + \eP(B_n')\notag\\
        \le & C\nth{\mathbf{1}(\overline{A_n}\cap \overline{B_n'})\max_{i,l\in[p]}\sup_{t\in\cT}\sqrt{nb_{i,l}}|\tilde \rho_{i,l}(t) - \rho_{i,l}(t)||\tilde \Gamma_{i,l}(t)-\Gamma_{i,l}(t)|}_{q/2}^{q/2}\delta^{-q/2} + \eP(A_n) + \eP(B_n')\notag\\
        := & CI_n\delta^{-q/2} + \eP(A_n) + \eP(B_n'),\label{eq:plugineq1}
    \end{align}
    where $C$ is an absolute constant, and $I_n$ is defined in an obvious way.

    In the following we evaluate $I_n$. By Cauchy-Schwarz inequality, \begin{align}\label{eq:pfD4eq1}
        I_n\le & \nth{\mathbf{1}(\overline{A_n}\cap \overline{B_n'})\max_{i,l\in[p]}\sup_{t\in\cT}\sqrt{nb_{i,l}}|\tilde \rho_{i,l}(t) - \rho_{i,l}(t)|}_{q}^{q/2}\nth{\mathbf{1}(\overline{A_n}\cap \overline{B_n'})\max_{i,l\in[p]}\sup_{t\in\cT}|\tilde \Gamma_{i,l}(t)-\Gamma_{i,l}(t)|}_q^{q/2}\notag\\
        := & I_n^{(1)}I_n^{(2)}.
    \end{align}
    On event $\overline{B_n'}$, if $n$ is large enough, there exists a constant $C$ satisfying $\inf_{t\in\cT}\min_{i,l\in[p]}\tilde\Gamma_{i,l}(t) > C$. This gives there exist another positive constant $C$, such that $|\tilde{\Gamma}_{i,l}(t)-\Gamma_{i,l}(t)|\le C\ath{\tilde{\Gamma}_{i,l}^2(t)-\Gamma_{i,l}^2(t)}$. So using Proposition 2 of \cite{bai2025time}, we have \begin{align}\label{eq:pfD4eq2}
        I_n^{(2)} \le C\nth{\max_{i,l\in[p]}\sup_{t\in\cT}|\tilde \Gamma_{i,l}^2(t)-\Gamma_{i,l}^2(t)|}_q^{q/2} = O\sth{p(g_n')^{q/2}}.
    \end{align}
    For the evaluation of $I_n^{(1)}$, recall the definition of $\vartheta_{i,l}(t), \Xi_{j,i,l}$ as in \eqref{eq8}, and $\cT_n=\{\nb,\dots,n-\nb\}$. From Lemma \ref{lm3}(\romannumeral 1), we have  $\sup_{i,l\in[p]}\sup_{j\in\cT_n}\nth{\vartheta_{i,l}(t_j)}_q = O(1)$. Using \eqref{pfthm1eq2}, \eqref{pfthm1eq3} and Minkowski inequality, we have \begin{align*}
      &\nth{\mathbf{1}(\overline{A_n}\cap \overline{B_n'})\max_{i,l\in[p]}\sup_{t\in\cT}\sqrt{nb_{i,l}}|\tilde \rho_{i,l}(t) - \rho_{i,l}(t)|}_{q} \le \nth{\sup_{i,l\in[p]}\sup_{j\in\cT_n}|\vartheta_{i,l}(t_j)|}_{q} +O\sth{c_n+p^{2/q}(nb)^{-1}}\\
      & = O\sth{c_n+p^{2/q}(nb)^{-1} + (p^2/b)^{1/q}} = O\sth{(p^2/b)^{1/q}}.
    \end{align*}
    The last line comes from the fact that $c_n\rightarrow 0, p^{2/q}(nb)^{-1}\rightarrow 0$ from bandwidth conditions in Assumption \ref{Ass:bdrate}. So we have $I_n^{(1)} = O(pb^{-1/2})$. By \eqref{eq:pfD4eq1} and \eqref{eq:pfD4eq2} we have $I_n = O\sth{p^2b^{-1/2}(g_n')^{q/2}}$. Recall that we have taken $q_n = \sth{g_n(np^2)^{1/q}}^{-1/(q+2)}$. By the bound of $I_n$, $\eP(A_n) = O(q_n^{-q}) = O\qth{\sth{g_n(np^2)^{1/q}}^{q/(q+2)}}, \eP(B_n') = O\sth{\pth{q_n'}^{-q}} = O\sth{\pth{\sqrt{nb}c_n}^{q/(q+1)}}$, and \eqref{eq:plugineq1} we finish the proof of Lemma \ref{lem:plugingap}.

    \subsection{Lemma \ref{lem:pvalE} and its Proof}

    Recall that $\cI_{i,l} = \{t \in\cT:(i,l)\in\cH_0(t)\}$ denote the set of time points at which the null hypothesis
$H_{i,l,t}^0$ is true. Its interior region, after removing a boundary
band of width $b = \max_{(i,l)\in\cH} b_{i,l}$, is defined by $\cI_{i,l}' = \{t: [t-b,t+b] \subset \cI_{i,l}\}$. Recall the definition of $A_{\Delta_n}$ in \eqref{eq:Asigman}. 

    \begin{lemma}\label{lem:pvalE}
    Assume all conditions in Theorem \ref{thm:pvalp} hold. Then we have \begin{align*}
        \eP\sth{ \inf_{t\in\cI_{i,l}'}P_{i,l}(t)\le x, A_{\Delta_n}} \le \eP\sth{\inf_{t\in\cI_{i,l}'}P_{i,l}(t)\le x \mid A_{\Delta_n}} \le \Delta_n + x + 1/B.
    \end{align*}
    \end{lemma}

    \textbf{Proof.} Recall the definition of $T_{i,l}(t),P_{i,l}(t)$ in Algorithm \ref{alg1}. Let $T_{i,l}^*(t)$ be an identically distributed and independent copy of centered version of $T_{i,l}(t)$, which is $(nb_{i,l})^{1/2}|\tilde \rho_{i,l}(t) - \rho_{i,l}(t)|/\tilde{\Gamma}_{i,l}(t)$. For $P_{i,l}(t)$ we have the following decomposition $P_{i,l}(t) = \frac{1}{B}\sum_{k=1}^B  \mathbf{1}\sth{T_{i,l}(t) < Z_{\bt,i,l}^{(k)}}$. The $P$-value satisfies $\inf_{t\in\cI_{i,l}'} P_{i,l}(t) \ge \frac{1}{B}\sum_{k=1}^B  \mathbf{1}\sth{\sup_{t\in\cI_{i,l}'}T_{i,l}(t) < Z_{\bt,i,l}^{(k)}}$.
    Given $\cF_n$, the terms in the summation are independent. 
    Therefore, \begin{align}
        &\eP\sth{\inf_{t\in\cI_{i,l}'}P_{i,l}(t) \le x\mid A_{\Delta_n}} \notag\\
        \le &\eE\sth{\eP\pth{\mathrm{Binom}\qth{\eP\sth{\sup_{t\in\cI_{i,l}'}T_{i,l}(t) < Z_{\bt,i,l} },B}\le \lceil xB\rceil\mid\cF_n}\mid A_{\Delta_n}}\notag\\
        \le & \eE\qth{\eP\sth{\mathrm{Binom}\pth{\qth{\eP\sth{\sup_{t\in\cI_{i,l}'}T_{i,l}(t) < \sup_{t\in\cT}T_{i,l}^*(t) \mid\cF_n}-\Delta_n} \vee 0,B}\le \lceil xB\rceil\mid\cF_n}}\notag\\
        \le & \eE\qth{\eP\sth{\mathrm{Binom}\pth{\qth{\eP\sth{\sup_{t\in\cI_{i,l}'}T_{i,l}(t) < \sup_{t\in\cI_{i,l}'}T_{i,l}^*(t) \mid\cF_n}-\Delta_n} \vee 0,B}\le \lceil xB\rceil\mid\cF_n}}.\label{pflmB5eq1}
    \end{align}
    Since $\sup_{t\in\cI_{i,l}'}T_{i,l}^*(t)$ has the same distribution as $\sup_{t\in\cI_{i,l}'}T_{i,l}(t)$, by Lemma \ref{lem:Cop}, $$\eP\qth{\eP\sth{\sup_{t\in\cI_{i,l}'}T_{i,l}(t) < \sup_{t\in\cI_{i,l}'}T_{i,l}^*(t)\mid\cF_n} \le a} \ge a$$ for $a\in(0,1)$. So by \eqref{pflmB5eq1},
    \begin{align*}
    &\eE\qth{\eP\sth{\mathrm{Binom}\pth{\qth{\eP\sth{\sup_{t\in\cI_{i,l}'}T_{i,l}(t)\le \sup_{t\in\cI_{i,l}'}T_{i,l}^*(t) \mid\cF_n}-\Delta_n} \vee 0,B}\le \lceil xB\rceil\mid\cF_n}}\\
        = & \Delta_n + \int_0^{1-\Delta_n} \sum_{k=0}^{\lceil xB\rceil} C^B_{k}p^k(1-p)^{B-k}\mathrm{d}p\le \Delta_n + \int_0^{1} \sum_{k=0}^{\lceil xB\rceil} C^B_{k}p^k(1-p)^{B-k}\mathrm{d}p \notag\\
        = & \Delta_n + \sum_{k=0}^{\lceil xB\rceil} \frac{B!}{k!(B-k)!}\frac{k!(B-k)!}{(B+1)!} \le \Delta_n + \frac{xB+1}{B+1} \le \Delta_n + x + 1/B.
    \end{align*}

    \subsection{Lemma \ref{lem:Cop} and its Proof}

    \begin{lemma}\label{lem:Cop}
        Assume $X, X'$ are independent and identically distributed random variables with distribution function $F(x)$. Then we have $\eP\sth{\eP(X < X'|X) \le a} \ge a$ for all $a\in (0,1)$.
    \end{lemma}

    \textbf{Proof.} Let $F^{-1}(a) = \inf\{x\in\eR:F(x) \ge a\}$ for $a\in(0,1)$. Then we have \begin{align*}
        \eP\sth{\eP(X < X'|X) \le a} &= \eP\sth{1-F(X) \le a} = \eP\sth{X \ge F^{-1}(1-a)}\\
        & = 1- \eP\sth{X' < F^{-1}(1-a)} = 1- \sup_{y < F^{-1}(1-a)} F(y).
    \end{align*}
    For any $y<F^{-1}(1-a)$, from the definition of $F^{-1}$, we have $F(y) < 1-a$. This gives $\sup_{y < F^{-1}(1-a)} F(y) \le 1-a$, and thus $\eP\sth{\eP(X < X'|X) \le a} \ge a$.

\setcounter{section}{2}
\section{Additional Simulations}\label{app:SecC}

\subsection{Baseline Comparison}\label{app:SecC1}

We compare our method to the moving-window approach (denoted as MW) described in Section~3.9 of \cite{masuda2025introduction}. This method selects a window length $w$ and, at time $t = j/n$, computes sample correlations using data from the time interval $[j-w, j+w]$. Correlation is detected at time $t$ if the sample correlation exceeds a chosen threshold $\tau$. As pointed out in \cite{masuda2025introduction}, there is no consensus on an optimal method for selecting $\tau$. For brevity, we report this moving-window comparison only for Cases~1--2 with Gaussian innovations. The results in Table~\ref{tab:mw} show that, depending on $\tau$, the method can either produce an excessively high AuFDP or be overly conservative.

\begin{table}[htbp]
    \centering
    \caption{Moving-window baseline results for Cases~1--2 with selected thresholds, with $r_n = 2(\log n)^{1/2}$.}
    \label{tab:mw}
    \begin{tabular}{cccccc}
        \toprule
        \multirow{2}{*}{$p$} & \multirow{2}{*}{$n$} & \multicolumn{2}{c}{$\mathrm{AuFDP}_{r_n}$} & \multicolumn{2}{c}{Average FNP} \\
        \cmidrule(lr){3-4} \cmidrule(lr){5-6}
        & & $\tau = 0.1$ & $\tau = 0.3$ & $\tau = 0.1$ & $\tau = 0.3$ \\
        \midrule
        \multirow{3}{*}{6} 
            & 600  & 0.300 & 0.005 & 0.000 & 0.062 \\
            & 900  & 0.264 & 0.000 & 0.000 & 0.047 \\
            & 1500 & 0.221 & 0.000 & 0.000 & 0.026 \\
        \midrule
        \multirow{3}{*}{9} 
            & 600  & 0.414 & 0.017 & 0.000 & 0.061 \\
            & 900  & 0.368 & 0.000 & 0.000 & 0.045 \\
            & 1500 & 0.291 & 0.000 & 0.000 & 0.027 \\
        \bottomrule
    \end{tabular}
\end{table}

We further compare Algorithm~\ref{alg1} with the multiple-testing procedure of \cite{cai2016large} (referred to as MT). The original MT procedure was developed for zero-mean stationary time series. To make it applicable in the present locally stationary setting, we implement a moving-window version: for a given bandwidth \(b\), the MT procedure is applied at time \(t\) using only observations with \(j/n \in (t-b/2,\, t+b/2)\). To ensure a fair comparison, we set the mean functions to zero when applying both the MT method and Algorithm~\ref{alg1}. For brevity, we also report the MT comparison only for Cases~1--2 with Gaussian innovations.

The results, summarized in Table~\ref{tab:compmt}, reveal several key limitations of the MT method compared to Algorithm~\ref{alg1}. First, the MT method exhibits a thresholding behavior (see Remark 1 of \cite{cai2016large}) and fails to respond to small nominal $\alpha$ levels, yielding identical $\mathrm{AuFDP}$ and $\mathrm{AuFNP}$ values across $\alpha = 0.05,$ and $0.10$. Furthermore, its $\mathrm{AuFDP}$ tends to rise with $n$ and exceeds the theoretical upper bounds established in our analysis. In contrast, our proposed method maintains stable AuFDP control across all tested $\alpha$ levels and sample sizes while demonstrating a consistent reduction in AuFNP as $n$ grows.

\begin{table}[htbp]
    \centering
    \caption{Comparison of Algorithm~\ref{alg1} with the MT method for Cases~1--2 across different levels of $\alpha$, with $r_n = 2(\log n)^{1/2}$.}
    \label{tab:compmt}
    \begin{tabular}{ccccccc}
        \toprule
        \multirow{2}{*}{Method} & \multirow{2}{*}{$p$} & \multirow{2}{*}{$n$} & \multicolumn{2}{c}{$\mathrm{AuFDP}_{r_n}$} & \multicolumn{2}{c}{Average FNP} \\
        \cmidrule(lr){4-5} \cmidrule(lr){6-7}
        & & & $\alpha = 0.05$ & $\alpha = 0.10$ & $\alpha = 0.05$ & $\alpha = 0.10$ \\
        \midrule
        \multirow{6}{*}{Alg \ref{alg1}} & \multirow{3}{*}{6} 
              & 600  & 0.040 & 0.058 & 0.084 & 0.062 \\
            & & 900  & 0.025 & 0.041 & 0.031 & 0.022 \\
            & & 1500 & 0.030 & 0.044 & 0.005 & 0.003 \\
        \cmidrule(lr){2-7}
            & \multirow{3}{*}{9} 
              & 600  & 0.046 & 0.060 & 0.107 & 0.081 \\
            & & 900  & 0.043 & 0.056 & 0.053 & 0.036 \\
            & & 1500 & 0.043 & 0.059 & 0.008 & 0.004 \\
        \midrule
        \multirow{6}{*}{MT} & \multirow{3}{*}{6} 
              & 600  & 0.088 & 0.088 & 0.006 & 0.006 \\
            & & 900  & 0.088 & 0.088 & 0.000 & 0.000 \\
            & & 1500 & 0.100 & 0.100 & 0.000 & 0.000 \\
        \cmidrule(lr){2-7}
            & \multirow{3}{*}{9} 
              & 600  & 0.082 & 0.082 & 0.011 & 0.011 \\
            & & 900  & 0.082 & 0.082 & 0.001 & 0.001 \\
            & & 1500 & 0.086 & 0.086 & 0.000 & 0.000 \\
        \bottomrule
    \end{tabular}
\end{table}

\subsection{Sensitivity Analysis}\label{app:SecC2}

In this section, we assess the robustness of Algorithm~\ref{alg1} with respect to the choice of tuning parameters. For brevity, we report sensitivity results only for Cases~1--2 under Gaussian innovations.

First, we consider perturbations of $\pm 10\%$ to the bandwidths $b_{i,l}$ selected via the data-driven procedure described in Section~\ref{sec.3.4}. Second, we evaluate the sensitivity to the lag parameter \(h\). Our main results use \(h=\lceil h^*\log n\rceil\) with \(h^*=1\), while the sensitivity to $h$ is assessed by considering \(\pm 20\%\) perturbations, namely \(h^*=0.8\) and \(h^*=1.2\).

Tables~\ref{tab:sensb} and~\ref{tab:sensh} report the sensitivity results, showing that Algorithm~\ref{alg1} is robust to perturbations of both the bandwidth and the lag parameter.

\begin{table}[htbp]
    \centering
    \caption{Sensitivity analysis for bandwidth perturbations in Cases~1--2 across different levels of $\alpha$, with $r_n = 2(\log n)^{1/2}$.}
    \label{tab:sensb}
    \small 
    \begin{tabular}{ccccccc}
        \toprule
        \multirow{2}{*}{Perturbation} & \multirow{2}{*}{$p$} & \multirow{2}{*}{$n$} & \multicolumn{2}{c}{$\mathrm{AuFDP}_{r_n}$} & \multicolumn{2}{c}{Average FNP} \\
        \cmidrule(lr){4-5} \cmidrule(lr){6-7}
        & & & $\alpha = 0.05$ & $\alpha = 0.10$ & $\alpha = 0.05$ & $\alpha = 0.10$ \\
        \midrule
        \multirow{6}{*}{-0.1} & \multirow{3}{*}{6} 
              & 600  & 0.024 & 0.036 & 0.173 & 0.138 \\
            & & 900  & 0.015 & 0.021 & 0.091 & 0.064 \\
            & & 1500 & 0.018 & 0.025 & 0.020 & 0.012 \\
        \cmidrule(lr){2-7}
            & \multirow{3}{*}{9} 
              & 600  & 0.027 & 0.036 & 0.207 & 0.170 \\
            & & 900  & 0.021 & 0.033 & 0.128 & 0.094 \\
            & & 1500 & 0.023 & 0.035 & 0.034 & 0.020 \\
        \midrule
        \multirow{6}{*}{0.0} & \multirow{3}{*}{6} 
              & 600  & 0.037 & 0.057 & 0.087 & 0.066 \\
            & & 900  & 0.024 & 0.043 & 0.032 & 0.024 \\
            & & 1500 & 0.031 & 0.042 & 0.005 & 0.003 \\
        \cmidrule(lr){2-7}
            & \multirow{3}{*}{9} 
              & 600  & 0.045 & 0.057 & 0.110 & 0.085 \\
            & & 900  & 0.041 & 0.056 & 0.055 & 0.037 \\
            & & 1500 & 0.043 & 0.060 & 0.008 & 0.004 \\
        \midrule
        \multirow{6}{*}{0.1} & \multirow{3}{*}{6} 
              & 600  & 0.057 & 0.077 & 0.041 & 0.031 \\
            & & 900  & 0.044 & 0.065 & 0.014 & 0.009 \\
            & & 1500 & 0.046 & 0.062 & 0.002 & 0.001 \\
        \cmidrule(lr){2-7}
            & \multirow{3}{*}{9} 
              & 600  & 0.067 & 0.083 & 0.053 & 0.037 \\
            & & 900  & 0.064 & 0.080 & 0.021 & 0.013 \\
            & & 1500 & 0.062 & 0.081 & 0.002 & 0.001 \\
        \bottomrule
    \end{tabular}
\end{table}

\begin{table}[htbp]
    \centering
    \caption{Sensitivity analysis for varying $h^*$ in Cases~1--2 across different levels of $\alpha$, with $r_n = 2(\log n)^{1/2}$.}
    \label{tab:sensh}
    \small
    \begin{tabular}{ccccccc}
        \toprule
        \multirow{2}{*}{$h^*$} & \multirow{2}{*}{$p$} & \multirow{2}{*}{$n$} & \multicolumn{2}{c}{$\mathrm{AuFDP}_{r_n}$} & \multicolumn{2}{c}{Average FNP} \\
        \cmidrule(lr){4-5} \cmidrule(lr){6-7}
        & & & $\alpha = 0.05$ & $\alpha = 0.10$ & $\alpha = 0.05$ & $\alpha = 0.10$ \\
        \midrule
        \multirow{6}{*}{0.8} & \multirow{3}{*}{6} 
              & 600  & 0.038 & 0.055 & 0.093 & 0.071 \\
            & & 900  & 0.033 & 0.046 & 0.037 & 0.026 \\
            & & 1500 & 0.023 & 0.034 & 0.008 & 0.004 \\
        \cmidrule(lr){2-7}
            & \multirow{3}{*}{9} 
              & 600  & 0.053 & 0.067 & 0.113 & 0.088 \\
            & & 900  & 0.036 & 0.051 & 0.053 & 0.036 \\
            & & 1500 & 0.034 & 0.046 & 0.015 & 0.009 \\
        \midrule
        \multirow{6}{*}{1.0} & \multirow{3}{*}{6} 
              & 600  & 0.037 & 0.057 & 0.087 & 0.066 \\
            & & 900  & 0.024 & 0.043 & 0.032 & 0.024 \\
            & & 1500 & 0.031 & 0.042 & 0.005 & 0.003 \\
        \cmidrule(lr){2-7}
            & \multirow{3}{*}{9} 
              & 600  & 0.045 & 0.057 & 0.110 & 0.085 \\
            & & 900  & 0.041 & 0.056 & 0.055 & 0.037 \\
            & & 1500 & 0.043 & 0.060 & 0.008 & 0.004 \\
        \midrule
        \multirow{6}{*}{1.2} & \multirow{3}{*}{6} 
              & 600  & 0.046 & 0.060 & 0.087 & 0.066 \\
            & & 900  & 0.027 & 0.043 & 0.036 & 0.026 \\
            & & 1500 & 0.028 & 0.039 & 0.007 & 0.005 \\
        \cmidrule(lr){2-7}
            & \multirow{3}{*}{9} 
              & 600  & 0.054 & 0.073 & 0.110 & 0.086 \\
            & & 900  & 0.044 & 0.062 & 0.057 & 0.040 \\
            & & 1500 & 0.048 & 0.058 & 0.013 & 0.008 \\
        \bottomrule
    \end{tabular}
\end{table}

\end{document}